\documentclass{sig-alternate-05-2015}

\usepackage{graphicx}
\usepackage{hyperref}
\usepackage{amsmath}
\usepackage{cite}
\usepackage{subcaption}
\usepackage{tabu}
\usepackage[font=small,format=hang,parskip=1pt]{caption}

\usepackage{amsthm}
\usepackage{float}
\usepackage{algorithm}

\usepackage{algorithmic}
\usepackage[usenames,dvipsnames]{xcolor}

\DeclareCaptionLabelFormat{noname}{#2}

\newlength\myindent
\usepackage{url}
\urldef{\mailsb}\path|{A.Shaghaghi, Sanjay.Jha}@unsw.edu.au|
\urldef{\mailsa}\path|Dali.Kaafar@data61.csiro.au |

\usepackage{pgfplots}
\usepackage{etoolbox}
\usepackage[normalem]{ulem}
\newcommand\redout{\bgroup\markoverwith
{\textcolor{red}{\rule[.5ex]{2pt}{0.4pt}}}\ULon}
\usepackage{xcolor}

\newcommand*{\affmark}[1][*]{\textsuperscript{#1}}

\begin{document}
\CopyrightYear{2017} 
\setcopyright{acmcopyright}
\conferenceinfo{ASIA CCS '17,}{April 02-06, 2017, Abu Dhabi, United Arab Emirates}
\isbn{978-1-4503-4944-4/17/04}\acmPrice{\$15.00}
\doi{http://dx.doi.org/10.1145/3052973.3053039}

\title{WedgeTail: An Intrusion Prevention System for the Data Plane of Software Defined Networks}

\numberofauthors{1} %  in this sample file, there are a *total*

\author{
\alignauthor
Arash Shaghaghi\affmark[1,2], Mohamed Ali Kaafar\affmark[2] and Sanjay Jha\affmark[1]\\
       \affaddr{\affmark[1]School of Computer Science and Engineering, The University of New South Wales (UNSW), Australia}\\
       \affaddr{\{a.shaghaghi, sanjay.jha\}@unsw.edu.au}\\
       \affaddr{\affmark[2]Data61, CSIRO, Australia}\\
       \affaddr{\{dali.kaafar\}@data61.csiro.au}\\
}

% There's nothing stopping you putting the seventh, eighth, etc.
% author on the opening page (as the 'third row') but we ask,
% for aesthetic reasons that you place these 'additional authors'
% in the \additional authors block, viz.

% Just remember to make sure that the TOTAL number of authors
% is the number that will appear on the first page PLUS the
% number that will appear in the \additionalauthors section.

%%%%%%%%%%%%%%%%%%%%%%%%%%%%%%%%%%%%%%%%%%%%%%%%%%%%%%%%%%%%%%%%%%%%%%%%%%%
% Abstract.
%%%%%%%%%%%%%%%%%%%%%%%%%%%%%%%%%%%%%%%%%%%%%%%%%%%%%%%%%%%%%%%%%%%%%%%%%%%

\maketitle
\begin{abstract}
Networks are vulnerable to disruptions caused by malicious forwarding devices. The situation is likely to worsen in Software Defined Networks (SDNs) with the incompatibility of existing solutions, use of programmable soft switches and the potential of bringing down an entire network through compromised forwarding devices. In this paper, we present WedgeTail, an Intrusion Prevention System (IPS) designed to secure the SDN data plane. WedgeTail regards forwarding devices as points within a geometric space and stores the path packets take when traversing the network as trajectories. To be efficient, it prioritizes forwarding devices before inspection using an unsupervised trajectory-based sampling mechanism. For each of the forwarding device, WedgeTail computes the expected and actual trajectories of packets and `hunts' for any forwarding device not processing packets as expected. Compared to related work, WedgeTail is also capable of distinguishing between malicious actions such as packet drop and generation. Moreover, WedgeTail employs a radically different methodology that enables detecting threats autonomously. In fact, it has no reliance on pre-defined rules by an administrator and may be easily imported to protect SDN networks with different setups, forwarding devices, and controllers. We have evaluated WedgeTail in simulated environments, and it has been capable of detecting and responding to all implanted malicious forwarding devices within a reasonable time-frame.  We report on the design, implementation, and evaluation of WedgeTail in this manuscript.   
\end{abstract}


%
%  Use this command to print the description
%
\printccsdesc

% We no longer use \terms command


\section{Introduction} \label{sec:introduction}
An attacker may compromise a network forwarding device by exploiting its software or hardware vulnerabilities. Compromised forwarding devices may be then used to drop or slow down, clone or deviate, inject or forge network traffic to launch attacks targeting the network operator and its users. As discussed in \cite{kloti2013openflow, kreutz2013towards}, compromised forwarding devices may even grant an attacker the capability to wrest control of an entire Software Defined Network (SDN). This paper looks at the specific problem of protecting SDNs from malicious forwarding devices by determining if the traffic forwarding function of the switch itself is secure. \par

Securing the network against malicious switches have not been the subject of many studies in SDN security research -- see \cite{ali2015survey, scott2015survey, kreutz2015software} for comprehensive surveys of SDN security. In fact, even with the latest proposals, there seems to be an oversight regarding the malicious forwarding devices that may exist in SDN data plane \cite{ghannam2016handling}. In general, the development of SDN security applications and controllers and real-time verification of network constraints, separately, have been the primary focus of SDN security literature (see\cite{ali2015survey, kreutz2015software} for surveys). However, no combination of these provides effective protection against compromised forwarding devices \cite{dhawan2015sphinx, ghannam2016handling, chao2016securing}. \par

Recently, a few proposals specifically look into the threats associated with malicious forwarding devices. However, these either suffer from a simplistic threat model (e.g.\cite{kamisinski2015flowmon}) or substantial processing overhead imposed to the network (e.g.\cite{AdrianSP, zhang2012shortmac, kim2014lightweight}). For example, cryptographic solutions such as \cite{kim2014lightweight} have been designed to enforce path compliance in the presence of strong adversaries, nevertheless, a universal deployment may be infeasible due to high overload required for per packet cryptographic operations, increased packet size, and etc. \par

SPHINX \cite{dhawan2015sphinx} is one of the solutions designed for securing the SDN data plane that does not assume forwarding devices are to be trusted. SPHINX detects and mitigates security attacks launched by malicious switches by abstracting the network operations with incremental flow graphs. It detects attacks as per the policies defined by the administrator and responds accordingly. SPHINX also checks for flow consistency throughout a flow path using a similarity index metric, where this metric must be similar for `good' switches on the path. 

We argue the following three factors as the main limitation of SPHINX. First, the system does not tolerate Byzantine forwarding faults. Therefore, SPHINX does not assume malicious forwarding device could behave arbitrarily and is not capable of distinguishing between malicious actions (e.g. packet drop and fabrication), and it cannot detect when a malicious forwarding device is delaying packets. Second, the detection mechanism mainly relies on the policies defined by an administrator to detect attacks. In fact, the flow-graph component does not validate forwarding device actions against the controller policies but only compared to their behavior over time -- hence, radical network configuration changes will lead to false positives. Moreover, the flow-graph feature requires that the majority of forwarding devices be trustworthy. Indeed, an alternative more robust solution will have to be independent of this assumption. Thirdly, SPHINX does not prioritize its inspection of forwarding devices. Arguably, an efficient solution should prioritize this task to improve detection performance.

Here, we introduce WedgeTail, a controller-agnostic Intrusion Prevention System (IPS) designed to `hunt' for forwarding devices failing to process packets as expected. WedgeTail regards packets as `random walkers' \cite{meloni2008scaling} in the network and analyzes packet movements as trajectories in a geometric space. By analyzing the expected and actual trajectories of  packets, our proposed solution is capable of automatically localizing malicious forwarding device and identifying the exact malicious behavior (e.g. packet drop, fabrication). WedgeTail response to threats can be programmed using administrator-defined policies. For example, an instant isolation policy may be customized such that initially, the potentially malicious device is instructed to reset all the flow rules and then, evaluated at various intervals by re-iterating the same packet(s) raising suspicion.

In order to make the scanning more efficient and increase the probability of finding malicious devices earlier, WedgeTail begins by prioritizing forwarding for inspection. We adopt Unsupervised Trajectory Sampling \cite{pelekis2010unsupervised} to cluster forwarding devices into scanning groups of varying priority depending on the cumulative frequency of occurrence in packet paths traversing the network. To retrieve the expected trajectories, WedgeTail intercepts the relevant OpenFlow messages exchanged between the control and data plane and maintains a virtual replica of the network. This virtual replica is processed by its integrated Header Space Analysis (HSA) \cite{kazemian2012header} component to calculate the expected packet trajectories. The actual packet trajectories are, however, computed by tracking a custom hash of the packet header. Alternatively, if NetSight \cite{NetSight} is deployed, WedgeTail queries for packet history to retrieve the packet trajectory. We briefly review \cite{NetSight} and \cite{kazemian2012header} in \S\ref{sec:background}. \par

The contributions of this work can be summarised as follows: 
      
\textbf{(a)} We define an advanced threat model for the security of SDN data plane that has not been considered up to now (\S\ref{sec:threatmodel}). In \S\ref{sec:WedgeTail}, we first discuss the main factors that exacerbate the protection of SDN networks against malicious forwarding devices. Thereafter, the requirements for an effective solution and the key insights behind our proposed solution is presented.

\textbf{(b)} In \S\ref{sec:targetidentification}, we present WedgeTail's target identification mechanism, where we detail how to retrieve the packet trajectories and analyze them to create scanning regions. 
%Our trajectory-based methodology is novel and may inspire developing a new set of security services in the future for SDNs. 
%Furthermore, WedgeTail's integration with NetSight highlights the practical advantages of employing this framework in Software Defined Networks.

%In \S\ref{sec:targetidentification}, we present our trajectory-based target identification technique. 

 %Accordingly and in order to secure an SDN network against such an adversay, Through \S\ref{sec:WedgeTail}, \S\ref{sec:response} and \S\ref{sec:implementation}, we discuss the design and implementation of WedgeTail, the first IPS for SDN data plane.

%\textbf{(c)} In \S\ref{sec:targetidentification}, we present two novel approaches through which, WedgeTail identifies and clusters targets for its inspection. {\color{red} \S\ref{subsub:openflowbased} is a solution to compute load at switch, port granularity in real-time using existing OpenFlow features and queue forwarding devices for inspection. In \S\ref{subsub:trajectorybased}, we introduce an unsupervised trajectory-based mechanism to cluster forwarding devices as per the frequency of occurrence in the packets routes. This is, in fact, the first attempt in the SDN literature to regard packets as `random walkers' \cite{meloni2008scaling} in the network and analyse packet movements as trajectories in a geometric space. We argue how the stripped architecture of SDN enables such methodologies that were complicated, and costly, to implement in traditional networks. [-- Arash: Change]}

\textbf{(c)} In \S\ref{sec:detection}, we present our proposed attack detection algorithms and localization logic. We also discuss how WedgeTail distinguishes between different malicious packet processing actions (e.g. packet replay and drop). WedgeTail's response engine and its capabilities are discussed in \S\ref{sec:response}.

\textbf{(d)} We discuss WedgeTail's implementation in \S\ref{sec:implementation}. Thereafter, in \S\ref{sec:evaluation}, we evaluate WedgeTail's performance and accuracy over three different simulated networks. We conclude the paper by comparing our solution with related work and outlining the future work (\S\ref{sec:discussion}).

%%%%%%%%%%%%%%%%%%%%%%%%%%%%%%%%%%%%%%%%%%%%%%%%%%%%%%%%%%%%%%%%%%%%%%%%%%%
%START OF SECTION: BACKGROUND
%%%%%%%%%%%%%%%%%%%%%%%%%%%%%%%%%%%%%%%%%%%%%%%%%%%%%%%%%%%%%%%%%%%%%%%%%%%
\section{Background} \label{sec:background}
\subsection{Header Space Analysis (HSA)} \label{sub:hsa}
Header Space Analysis (HSA) \cite{kazemian2012header} is a method for debugging network configuration. HSA deals with a L-bit packet header as L-dimensional space, and models all processes of routers and 
middle-boxes as transfer functions, which transform subspaces of the L-dimensional space to other subspaces. Therefore,  by analyzing forwarding rules of the network, HSA can calculate the path a packet traversing the network on a certain port will take. We have included an example usage of HSA and how it serves for predicting packet trajectories in \S\ref{sec:WedgeTail}.

\subsection{NetSight}
NetSight \cite{NetSight} is a network troubleshooting solution that allows SDN application to retrieve the packet history. \textit{netshark} is an example of tools built over this platform, which enables users to define and execute filters on the entire history of packets. With this tool, a network operator can also view the complete list of packet's properties at each hop, such as input port, output port, and packet header values. In \S\ref{sec:detection} we show how WedgeTail may inter-operate with NetSight to retrieve the actual packet trajectories. 

%%%%%%%%%%%%%%%%%%%%%%%%%%%%%%%%%%%%%%%%%%%%%%%%%%%%%%%%%%%%%%%%%%%%%%%%%%%
%START OF SECTION: The Key Idea
%%%%%%%%%%%%%%%%%%%%%%%%%%%%%%%%%%%%%%%%%%%%%%%%%%%%%%%%%%%%%%%%%%%%%%%%%%%

\section{Threat Model} \label{sec:threatmodel}
We assume a resourceful adversary who may have taken full control over one, or all, of the forwarding devices. This is, in fact, the strongest possible adversary that may exist at the SDN data plane, which to the best of our knowledge is not considered in the related work. 
For example, \cite{dhawan2015sphinx, kazemian2013real, kazemian2012header, khurshid2013veriflow, mai2011debugging}, assume all, or the majority, of the forwarding devices to be trustworthy. 
Interestingly, we have noticed an imprecise definition of adversary leading to oversights in SPHINX \cite{dhawan2015sphinx}, the closest work to ours. For instance, authors discuss an attack exhausting the TCAM memory of a switch that will cause a switch dropping packets over a period of time. As devastating as this may be, this device cannot be used to execute attacks requiring packet modification or misrouting. Here, we assume the following capabilities for the adversary: 
\begin{itemize}
	\item The attacker may drop, replay, misroute, delay even generate (includes both modify and fabricate) packets, in random or selective manner all or part of the traffic.
\end{itemize}
The above capabilities grant the adversary the capability to launch attacks against the network hosts, other forwarding devices or the control plane. For example, executing a Denial of Service (DoS) attack against the control plane by replaying or spoofing $Packet\_In$ messages. Note that detecting packet reordering is currently out of scope (\S\ref{sec:conclusion}. \par

We regard a forwarding device as `malicious' when both of the following properties are met: A) The device is not handling the network packets according to the rules specified by the control plane. B) The maliciousness is cloaked from basic troubleshooting tools. For example, the malicious device `correctly' responds to \textit{ping} or \textit{traceroute} probes while corrupting other packets.

Arguably, the above characteristics may also be witnessed with a misconfigured, or a faulty, forwarding device too. In fact, the differentiating factor between these is the underlying intentions and hardly their behavior or impact. Hence, for the purpose of this work, we expand the definition of a malicious forwarding device to encompass both faulty and misconfigured devices. This implies that the proposed solution could also be used to detect faulty and misconfigured forwarding devices which are functioning anomalously -- see Section \ref{sec:discussion}. \par

We make the following assumptions for WedgeTail to work:

1. The control plane itself and the defined policies are trustworthy and securely transferred to the data plane (e.g. using TLS protocol\cite{benton2013openflow}). There is an increasing body of literature aiming to achieve this, see \cite{scott2015survey, ali2015survey} for surveys. In other words, with SDN, the policy definition and enforcement points are separated in networks \cite{kreutz2015software} and here, we exclusively focus on the the Policy Enforcement Point (PEP). Hence, preventing incidents such as \cite{hunter2008pakistan} caused by erroneous administrator defined policies is out of scope.

2. Packet reordering and time behaviour\cite{mizrak2005fatih} are well-studied and proposed solutions are complementary to WedgeTail. This is also true regarding protocol-level attacks including TCP/IP and OSPF that can be addressed using existing solutions. In fact, WedgeTail is designed to detect forwarding devices failing to execute their main function and 
not to protect them from being compromised. In other words, the prevention refers to the automated triggering of pre-defined policies against identified threats.

3. The forwarding devices may lie about anything except their own identity -- similar assumption is also made in \cite{dhawan2015sphinx}.

\section{WedgeTail} \label{sec:WedgeTail}
As mentioned in \S\ref{sec:introduction}, securing SDN networks against malicious forwarding devices is challenging. In fact, similar to \cite{dhawan2015sphinx}, we also argue that the problem of protecting networks and their host against malicious forwarding devices is exacerbated in SDN context. We believe this due to five main reasons -- the first three factors are extracted from \cite{dhawan2015sphinx} with some minor amendments and additions. \par

First and foremost is the incompatibility of existing solutions to secure SDN. In fact, due to the removal of intelligence from the forwarding devices, the defense mechanisms used for traditional networks may no longer work. \cite{dhawan2015sphinx} postulates that for a comprehensive defense against traditional attacks either a fundamental redesign of OpenFlow \cite{mckeown2008openflow} protocol would be required, or we would need to patch the controller per each attack. \par

The second factor is the unverified and complete reliance of control plane on forwarding devices. An SDN controller relies on $PACKET\_IN$ messages for its view of the network, yet this is not securely authenticated nor verified. A malicious forwarding device may send forged spoofed messages to subvert the controller view of the network -- even with having TLS authentication in place. The same vulnerability enables a compromised forwarding device the capability to overload the controller with requests causing a Denial of Service (DoS) attack. \par

Third, securing programmable soft-switches such as Open vSwitches is more challenging compared to hardware equivalents. The former run atop of end host servers and are more susceptible to attacks compared with the hardware switches, which is harder for an attacker to physically access. \par

Our fourth argument is that the SDN security domain is a moving target with the protocols and standards undergoing constant changes. For example, several controllers have already been proposed with varying specifications, which are undergone constant updates. Hence, relying on the capabilities of one would limit practicality on another. The same argument is also valid on the OpenFlow \cite{OPenF} standard.

Fifth, motivated by performance advantages such as lower latency response to network events and better protocol standardization such as for encryption, MAC learning and codec control message (CCM) exchanges, proposals have been made to delegate more control to the SDN data plane \cite{onftr}. The increased authority improves the network's fault tolerance with the continuation of basic network operations under failing controllers. However, this increases the vulnerability of the network to traditional attacks and expands the range of attacks a compromised device may launch against the network.

Considering the aforementioned factors and the limitations of existing work such as SPHINX (see \S~\ref{sec:introduction}), we posit the `must-have' features of an effective solution against malicious forwarding devices to include: \\
1) Minimum reliance on pre-defined rules/policies for detection:  to be able to detect unknown attacks exploiting either hardware or software vulnerabilities of forwarding devices, \\
2) Ability to systematically and autonomously prioritize forwarding devices for inspection: to improve detection performance and success rate, \\
3) Capability to distinguish malicious forwarding actions and localize maliciousness: to avoid executing conflicting and redundant policies when responding to threats, \\
4) Programmability for responding to threats: to be able to customize response when detecting threats, and \\
5) Causing minimal disruption to the network performance when detecting and responding to threats: so the proposed solution is practical for real-world network deployment. \\
The rest of this paper describes how we address each of the aforementioned requirements in our proposed solution, WedgeTail.

  \begin{figure}
  \centering
  \includegraphics[height=2in, width=3.44in]{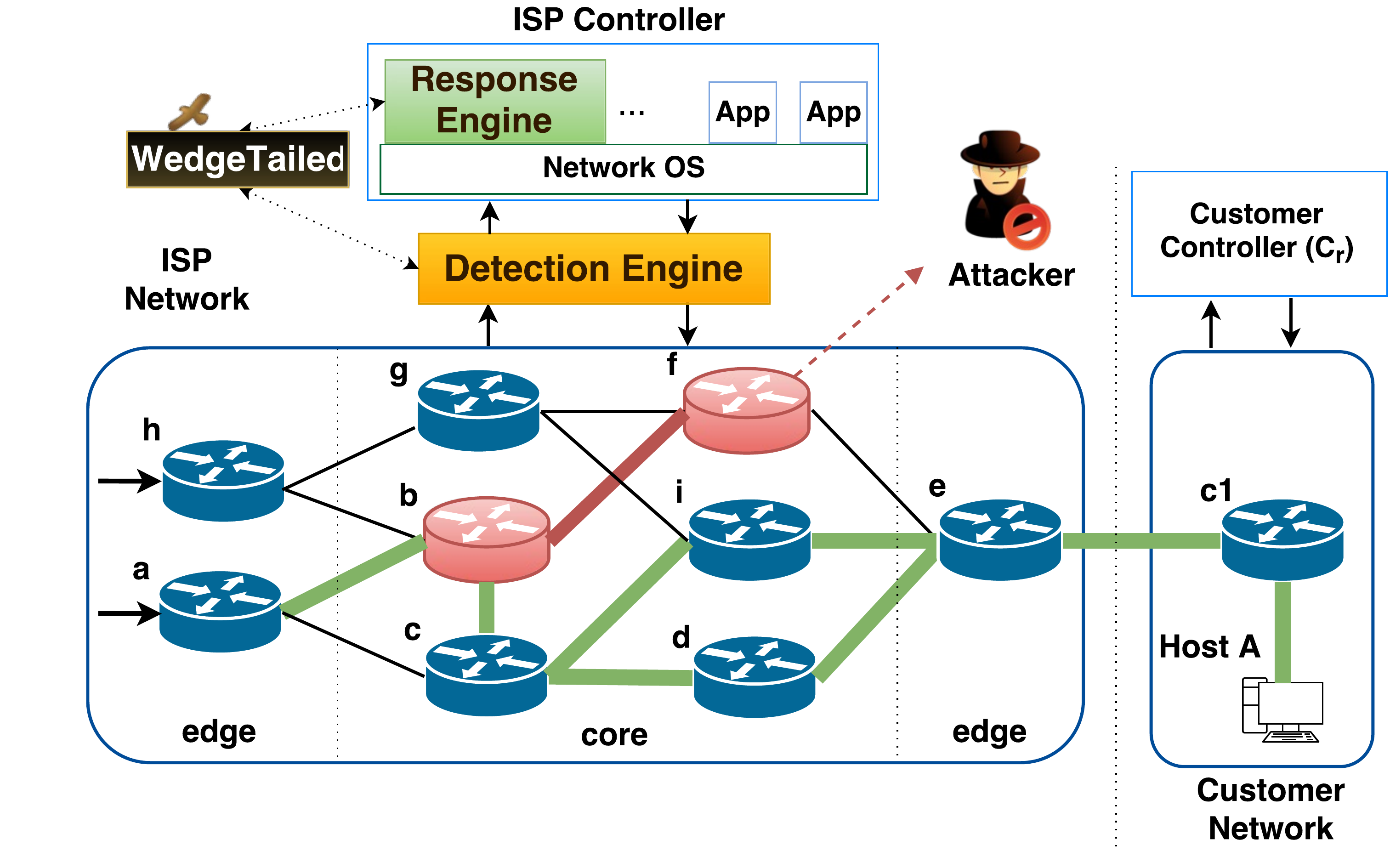}
  \caption{Forwarding devices in ISP Network. The red devices are misbehaving. The green links are expected paths.}
  \label{Figure1:keyidea}
  \end{figure}

\subsection{Overview}
As shown in Figure \ref{Figure1:keyidea}, WedgeTail is composed of two main parts namely, Detection Engine (\S\ref{sec:detection}) and Response Engine (\S\ref{sec:response}). The former listens to OpenFlow messages exchanged between the control and data plane and by doing so maintains a virtual replica of the network, which is used to compute the expected packet trajectories. The Response Engine, however, is placed as an application on top of the controller and submits policies to the network operating system, which makes the final decision on how, and whether, to apply them. \par

%entails Find-Target-Forwarding-Devices() and Scan-for-Attacks() functions listed in Algorithm \ref{algorithm1} and is responsible for identifying and clustering forwarding devices for inspection and detect maliciousness. The latter, however, includes the Isolate-Forwarding-Device() function, which submits recommendation policies to the network operating system, which makes the final decision on how, and whether, to apply them. 

Now, assume an ISP network with AARNet Setup traffic flow as shown in Figure \ref{Figure1:keyidea}. Using its integrated HSA component, WedgeTail retrieves 10010x10 $\cup$ 10011x10 as the header space of packets that can be sent from Forwarding Device  (FD) a to c1, or $FD(a) \rightarrow FD(c1)$, on $Port_i$. Composing the history attributes of the propagation graph it learns that the Expected Packet Trajectories between these two nodes is as follows:

%expected trajectory of packet $packet(i)$ 
%WedgeTail computes the expected trajectory of packet $packet(i)$, or $TR_{packet(i)}$, using its integrated Header Space Analysis (HSA) framework. In this case, WedgeTail retrieves the set of headers that ISP's $FD(a)$ can send to Customer's $FD(c1)$ on $Port_i$ is 10010x10 $\cup$ 10011x10. 

\noindent $FD(a) \rightarrow FD(b) \rightarrow FD(c) \rightarrow \{ FD(d)~OR~FD(i) \} \rightarrow FD(e) \rightarrow FD(c1)$ -- shown in green colour in Figure \ref{Figure1:keyidea}.

%The two green lines show trajectories for packet $packet(i)$ denoted as, flowing from ISP's forwarding device labeled $FD(a)$ to $FD(c1)$ in the customer network. As also shown in the figure, a $packet(i)$ may be routed differently due to networking configurations such as Quality of Service (QoS). WedgeTail comp

%Each of the forwarding devices en route of a packet, handle the incoming traffic according to their own flow table. We denote the flow table of $FD(i)$ as $FT_{FD(i)}$. The flow table rules are set by a controller $C$ in accordance with the respective network policy defined by the network administrator. \par
%Each of the forwarding devices en route of a packet, handle the incoming traffic according to their own flow table. We denote the flow table of $FD(i)$ as $FT_{FD(i)}$. The flow table rules are set by a controller $C$ in accordance with the respective network policy defined by a network administrator, which is shown as $P(C)_{ISP}$. \par

The main intuition behind WedgeTail is that whenever the \textit{Actual Packet Trajectories} are not a subset of the \textit{Expected Packet Trajectories}, one or more of the forwarding devices in the packet path may be malicious -- recall that in \S\ref{sec:threatmodel} `malicious' was extended to cover faulty and misconfigured too.
For instance, in Figure \ref{Figure1:keyidea}, the red colored trajectory is a non-allowed trajectory and
$FD(b)$ is malicious. 

Algorithm \ref{algorithm1} presents WedgeTail's workflow. On each run, WedgeTail inspects the whole network on a specific port -- out of the designated target ports. The detection engine entails Find-Target-Forwarding-Devices() and Scan-for-Attacks(), which provide input to the main protection engine function, Isolate-Forwarding-Device().

\begin{algorithm}[h]
    \begin{algorithmic}
    \caption{WedgeTail Detection and Response} \label{algorithm1}
    \STATE Response Policy $RP$
    \STATE Select $Port_i \in $ \{Port\}
    \STATE    \textbf{Find-Target-Forwarding-Devices}~($Port_i$)
    \STATE    Select $FD(i) \in$ \{Target Forwarding Devices\}
      \FORALL {$Port_{i} \in$ \{Port\}}
            \STATE \textbf{Scan-for-Attacks}~($FD(i)$)
                \IF {$FD(i)$ is `Malicious'}
                   \STATE \textbf{Isolate-Forwarding-Device}~(RP, $FD(i)$)
                \ENDIF
       \ENDFOR
    \end{algorithmic}
    \end{algorithm}
    
%%%%%%%%%%%%%%%%%%%%%%%%%%%%%%%%%%%%%%%%%%%%%%%%%%%%%%%%%%%%%%%%%%%%%%%%%%%
%START OF SECTION: Target Identification
%%%%%%%%%%%%%%%%%%%%%%%%%%%%%%%%%%%%%%%%%%%%%%%%%%%%%%%%%%%%%%%%%%%%%%%%%%%

\section{Target Identification} \label{sec:targetidentification}

\subsection{Trajectory Creation} \label{sub:packettracking} 

\textbf{Definition of Trajectory:}
We first define the notion of trajectory, denoted as $TR$ hereon, in the context of our work. A packet trajectory is the route a uniquely identifiable packet takes while traversing a network from one forwarding device to another. We consider different paths for the same packet as distinctive trajectories. In other words, a packet may be routed through various paths in respect to network configurations and condition on each iteration. For instance, as shown in Figure \ref{Figure1:keyidea}, a packet traversing through green line from $FD(a)$ to $FD(e)$ may be routed through $FD(i)$ or $FD(d)$ depending on the QoS requirements.
 However, multiple repetitions of the same path for the same packet is only regarded as one trajectory. 
\par
\textbf{Retrieving the Actual Packet Trajectories:}
     We propose two alternative solutions to retrieve the packet trajectories. As succinctly reviewed in \S\ref{sec:background}, NetSight is a recently proposed network troubleshooting solution that allows retrieving all the forwarding devices that a packet visited while traversing the network. Therefore, if NetSight was deployed in a network, a convenient approach would be to query for each packet header route and create the trajectories. This may be achieved by integrating a simple module for WedgeTail, which uses existing API provided by NetSight. Our preferred method to retrieves the actual trajectories is through NetSight. \par

An alternative approach would be for WedgeTail to run a deterministic hash function over the packet header and use this hash to track packet as it traverses the network (i.e. generating labels). The choice of an appropriate hash function would be crucial for this matter as is selecting the proper packet header values. To achieve this, we use the packet hashing function used in \cite{duffield2000trajectory}. We then pick packet header values such as source address, destination address from IP header and source port and destination port in TCP header. All the values used for the hashing are shown in Figure 3. 
Note that in practice the labels can be quite small (e.g., 20 bit) -- although the size of the packet labels depends on the specific situation. Evidently, in this case, the overhead to collect trajectory samples is small since the traffic that has to be collected from nodes only consists of such labels (plus some auxiliary information) \cite{duffield2000trajectory}.

An issue to consider is that in the unlikely case that collisions were to occur, WedgeTail's performance will not be affected. This is because such collisions will break the order of forwarding devices when retrieving trajectories and will result in invalid trajectories. Moreover, we envision the hashing-based solution to be used as an alternative where NetSight is not available and at most within small networks, where collisions are much less likely to occur.

\subsection{Scanning Zones} \label{sub:scanningzones}
\begin{figure}
  \centering
	\includegraphics[height=2in, width=3.5in]{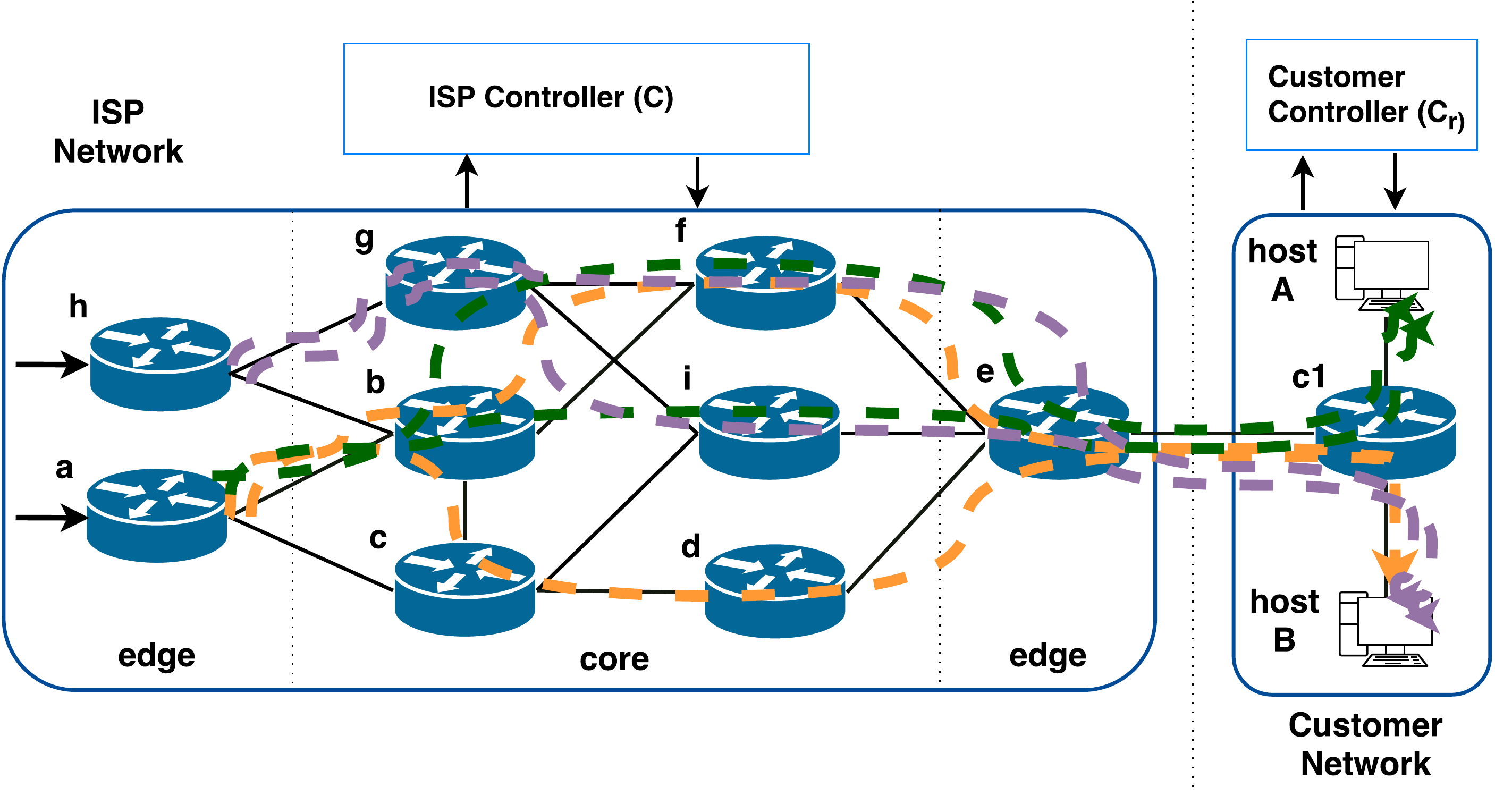}
  \caption{Forwarding devices in ISP network. The dotted lines represent traffic paths.}
  \label{Figure:Congestion}
\end{figure}

WedgeTail prioritizes forwarding devices for its inspection. The core idea is that the analysis has to begin from the forwarding devices that the majority of packets encounter while traversing the network. To identify these, WedgeTail keeps track of trajectories for all packets on all ports over time and identifies the most commonly involved forwarding devices by looking at the denser regions. For instance, looking at Figure \ref{Figure:Congestion} and the drawn trajectories, it is evident that $FD(b), FD(g)$ and $FD(f)$ are more commonly encountered by packets. Indeed, identifying these is much more complicated in a large network with a huge number of trajectories. Therefore, WedgeTail reduces this large set into a representative sample that encapsulates the most commonly visited forwarding devices. Formally, let $TR[FD(i), FD(j)]$ denote the set of all the trajectories traversing between $FD(i)$ and $FD(j)$ for all packets on all ports. Accordingly, define $\{TR(N)\}$ as the set of all trajectories in network N, or $\{TR(N)\} = \{TR[FD(i), FD(j)]\,|\,\forall (FD(i), FD(j)) \in N\}$. Denote $\{TR(N)\}'$ as a subset of $\{TR(N)\}$, which if inspected by WedgeTail without loss of generality results in detecting compromised forwarding devices. Indeed, such sampling is challenging due to the complexity of packet routing (e.g. lack of ordering, lack of compact representation). To automatically compute this, in an optimised and unsupervised way, WedgeTail adapts the Unsupervised Trajectory Sampling technique proposed in \cite{pelekis2010unsupervised}. \par

    The three-step Unsupervised Trajectory Sampling solution proposed in \cite{pelekis2010unsupervised} may be summarized as follows. The first step involves adopting a symbolic representation of trajectories to model all of the collected trajectories in an approximate way as vectors.  The symbolic representation is lossless in terms of mobility patterns and improves the speed of computation. Thereafter, on top of the representation, each trajectory is represented using a continuous function that implicitly describes the representativeness of each constituent part of it with respect to all of the collected trajectories. Symbolic Trajectory algorithm, or SyTra, is used to improve the initial representation of each trajectory by relaxing its vector representation. `The idea is to adopt a merging algorithm that identifies the maximal time period wherein the mobility pattern of each trajectory is preserved, while in this augmented period it presents uniform behavior in terms of representativeness' \cite{pelekis2010unsupervised}. In the third step, an automatic method for trajectory sampling, known as T-Sampling, based on the representativeness of the trajectories is used. T-Sampling takes into account not only the most (i.e., dense, frequent) but also the least representatives. This is an important aspect of this work, which makes it the best match for our requirements. In fact, alternative sampling techniques suffer from shortcomings that limit their application for WedgeTail. For example, \cite{andrienko2009visual, andrienko2009interactive} are explorative and supervised sampling techniques that assume a priori knowledge of the underlying trajectories. Alternatively, \cite{giannotti2007trajectory, lee2008trajectory} downsize the collection of trajectories and fail to select trajectories important for mobility patterns. \par
    Once the most commonly visited forwarding devices are extracted from the network they are allocated the highest priority of inspection and the remaining forwarding devices are assigned a lower priority for inspection. \par
	\begin{figure}
  \centering
	\includegraphics[height=1.4in, scale=0.5]{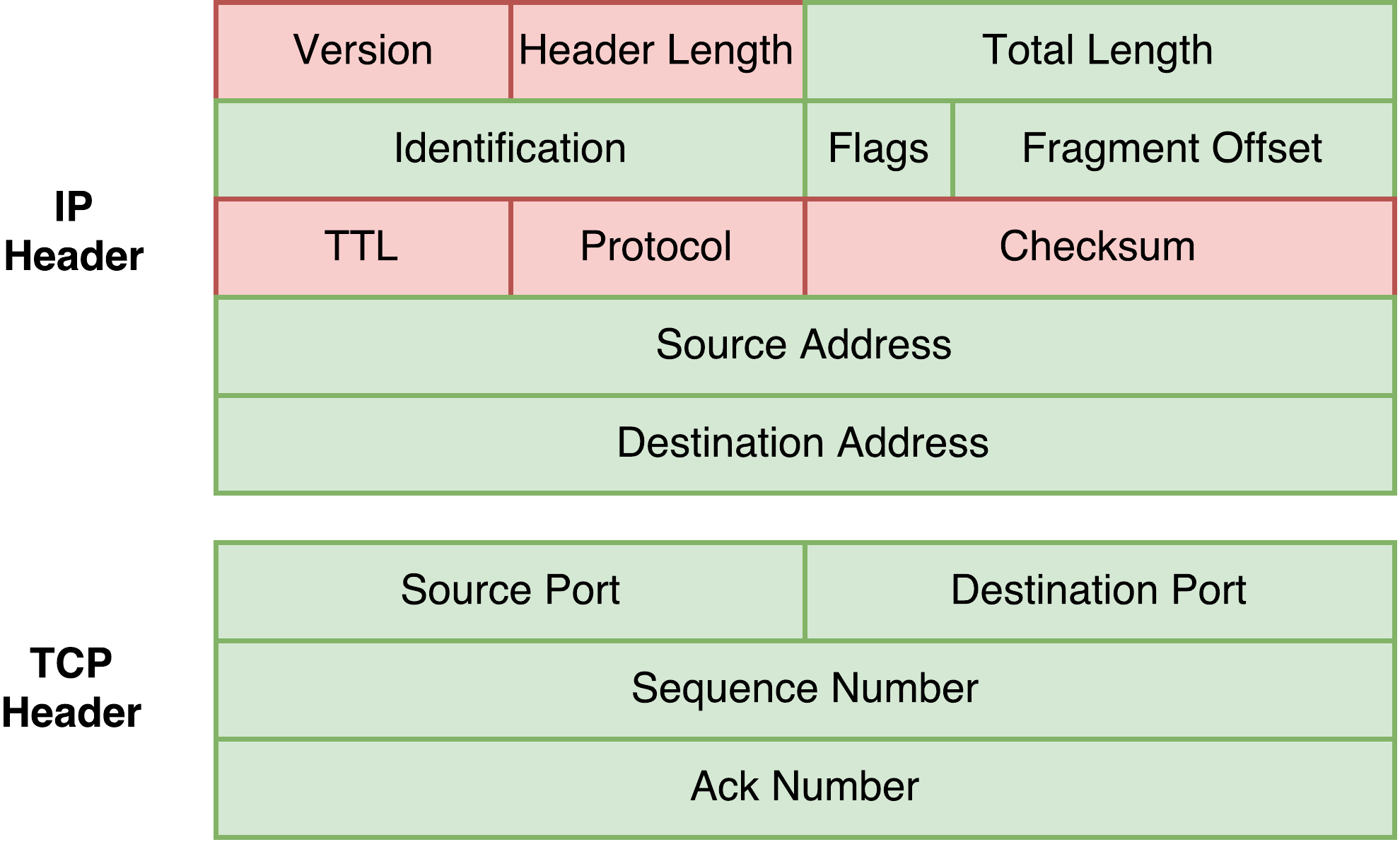}
  \caption{Packet header fields used for labeling.}
  \label{packetheaderfields}
\end{figure}	
%In traditional networks, network monitoring is regarded as an expensive task requiring additional hardware and/or software configuration. In this section, we present two mechanisms to compute load at \textit{(switch, port)} granularity in SDN. This is an important requirement for WedgeTail as it triggers the software attention for initiating its automated scan.  \par

%Although it is trivial to establish whether there exists a trajectory between two nodes in an arbitrary graph, counting the total number of them for general graphs is \#P. Hence, 

%%%%%%%%%%%%%%%%%%%%%%%%%%%%%%%%%%%%%%%%%%%%%%%%%%%%%%%%%%%%%%%%%%%%%%%%%%%
%START OF SECTION:Packet Tracking and Attack Detection
%%%%%%%%%%%%%%%%%%%%%%%%%%%%%%%%%%%%%%%%%%%%%%%%%%%%%%%%%%%%%%%%%%%%%%%%%%%

\section{Attack Detection} \label{sec:detection}

The main attack detection algorithm (\textit{Scan-for-Attacks}) is presented in Algorithm \ref{alg:attackdetection}. The algorithm takes as input both a target forwarding device and a port and returns a malicious node detailing its malicious action. First, a snapshot of all network forwarding device configurations is retrieved. Accordingly, the trajectories that a packet may take against each of the other forwarding devices, and the control plane, is computed --  Note that the packets required for creating the trajectories are chosen randomly in the control plane and cannot be known by an attacker to influence this process. Thereafter, the actual trajectories for the selected packets are retrieved using mechanisms discussed in \S\ref{sub:packettracking}. At this point, whenever the set of forwarding devices in the actual trajectory is not a subset of the expected trajectories, a malicious forwarding device is detected.

Formally, let $A$ denote the total ordered set of actual forwarding devices for a packet traversing from target $FD(i)$ to $FD(j)$ and $E$ the ordered set of expected forwarding devices for the same trajectory. If A $\not\subseteq$ E then $FD(i)$ is malicious. The comparison logic can be extended to differentiate between the four types of malicious actions (see \S\ref{sec:threatmodel}) as follows:

    \begin{algorithm}[h]
    \caption{Attack Detection Algorithm} \label{alg:attackdetection}
    \begin{algorithmic}
    \STATE Scan-for-Attacks($FD(i), Port P_{i}$) \{
    \STATE Status S = Check-State-Change();
    \STATE File F = Dataplane-Configrautions-Snapshot(S);
    \WHILE {Check-State-Change() ==  S}
    \STATE List L = F.ForwardingDevices() -- $FD(i)$
     \FORALL{$FD(j) \in$ L}
     \STATE Packet $Pck$;
     \STATE Trajectory Actual, Expected;
     \STATE Pck.Source() = $FD(i)$;
     \STATE Pck.Destination() = $FD(j)$;
     \STATE $Pck$ = Find-Packet(Pck.Source, \\ Pck.Destination);
     \STATE Expected = HSA-Trajectory($Pck$);
     \STATE Actual = Actual-Trajectory($Pck$);
     \IF {Actual $\neq$ Expected}
     \STATE Identify-Attack($FD(i), Port(i)$);
     \ENDIF 
    \ENDFOR
    \ENDWHILE
    \STATE \}
    \end{algorithmic}
    \end{algorithm}
    
Here, without loss of generality we assume, there exists only one valid trajectory between two forwarding devices. \par

\textbf{1. Packet Replay:} Occurs when a forwarding device sends a copy of the packet to a third destination as well as the intended destination. Figure \ref{Figure1:keyidea} shows a packet replication attack example, where $FD(b)$ replicates packets to $FD(f)$ which in turn an attacker may use to forward some, or all, of traffic to a machine under his control. A forwarding device that replays packets(s) enables an attacker to execute attacks such as surveillance and authentication attacks. \par

\textbf{Detection:} \textit{Let $FD(k)$ be a forwarding device other than $FD(i) and FD(j)$. $A'$ be the set of forwarding devices in the actual path excluding $FD(k)$, or $A - \{FD(k)\}$. If $ \exists FD(k) \in A: FD(k) \not\in E$ and $A' \subseteq E$ then WedgeTail detects a packet replay attack.}
 
\textbf{2. Packet Misrouting:} Occurs when a packet is diverged from the original destination and does not reach its intended destination. This may be used to launch an attack against network availability or as part of more complicated threats. For example, by forming a triangle routing and creating routing loop resulting in packet TTL value expiration the network congestion may result in a partial, or total, shutdown of the network. 

\textbf{Detection:} \textit{
Let $FD(k)$ be a forwarding device other than \{$FD(i), FD(j)$\} and $A'$ be the set of forwarding devices in the actual path excluding $FD(k)$, or $A - \{FD(k)\}$. If $ \exists FD(k) \in A: FD(k) \not\in E$ and $A' \not\subseteq E$ then WedgeTail detects a packet misrouting attack.
}

\textbf{3. Packet Dropping:} A compromised forwarding device that drops packets creates a black or gray hole in the network. In the former, it drops all the packets, and in the latter, it drops packets periodically or retransmission of packets or drops packets randomly. Packet dropping is used in attacks such Denial of Service (DoS) against network provider. WedgeTail detects packet dropping as follows:

\textbf{Detection:} \textit{
If $A \not\subseteq E$ and card(A) $<$ card(E).
}

\textbf{4. Packet Generation:} A compromised forwarding device may fabricate packets or modify existing ones. This may be used to mount attacks such as DoS. Such changes are detected by WedgeTail through its labeling mechanism. In other words, once any attribute used for labeling packets is changed, the label is changed, and the trajectory is undefined. WedgeTail detects packet generation as follows:

\textbf{Detection:} \textit{
    If $A \not\subseteq E$ and $E - A = E$.
}

 \textbf{5. Packet Delay:} Occurs when a forwarding device delays traffic and increases jitter. A packet delay is a threat against time-sensitive traffic \cite{ghannam2016handling}. A delay of TCP stream also causes spurious timeouts and unnecessary re-transmission, which severely threatens the TCP throughput \cite{zhu2005making}. 

\textbf{Detection:} \textit{
Let $T_e$ be the estimated time for $packet_i$ moving from $FD(i)$ and $FD(j)$ over a trajectory $\bar{\tau}$. Accordingly, let $T_a$ be the actual time that it took for this packet to traverse $\bar{\tau}$. Assume the maximum valid delay due to network congestion on this trajectory is $T_d$. If $\Delta T_{e,a} > T_d$ then there is a packet delay attack.
}

Note that the estimated time may be set to be the average time for all packets traversing that route or computed by sending simple \textit{ping} packets. The maximum valid congestion may be computed using \cite{rasley2015planck} or \cite{suh2014opensample}, where it is possible to achieve real-time congestion detection and measurement.
\subsection{Malicious Localization} 
As mentioned a trajectory is regarded as a total ordered set. Once one of the malicious actions is detected, it is possible to locate the associated forwarding device by comparing $A$ and $E$ (see previous section). Consider Figure \ref{Figure1:keyidea} and assume when inspecting $FD(a)$ we retrieve $E(\bar{\tau})$ and $A(\bar{\tau})$ as expected and actual trajectories between $FD(a)$ and $FD(e)$, respectively. 

\textbf{$A(\bar{\tau})$}: $FD(a) \rightarrow FD(b) \rightarrow FD(f) \rightarrow FD(e)$ equivalent to total ordered set  $E = \{FD(b), FD(f)$\}.

\textbf{$E(\bar{\tau})$}: $FD(a) \rightarrow FD(b) \rightarrow FD(c) \rightarrow FD(d) \rightarrow FD(e)$ equivalent to total ordered set $A = $\{$FD(b), FD(c), FD(d)$\}

In this case, by intersecting $E$ and $A$ we retrieve that \{$FD(b)$\} is the malicious node, where packet misrouting was initiated. The analysis is continued with $FD(c)$ and $FD(d)$ (or, $A - E$,) so that at the end of this process any forwarding device that my be malicious is identified. The same approach can be used for malicious actions 1, 3 and 4. To locate a forwarding device that is delaying packets however, we retrace time hop by hop in $A$ and compare with the relevant expected time.    

\begin{figure*}[!htb]
  \centering	
	\includegraphics[width=7.3in,scale=0.50,height=3in]{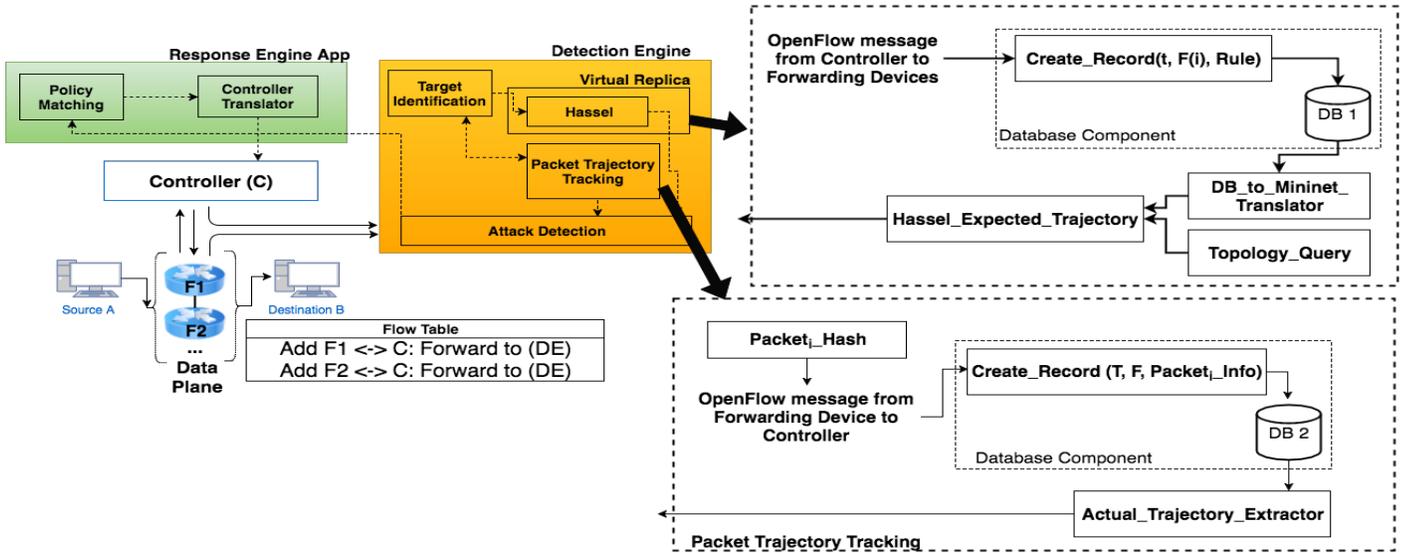}
  \caption{WedgeTail Architecture}
  \label{Figure4}
\end{figure*}

\subsection{Practical Considerations}\label{sub:practicalconsiderations}
Network congestion will result in packet drops and delays. Therefore, to minimize the number of false positives, WedgeTail has to estimate with a high accuracy the number of packets drops and delays associated with network congestion. Several solutions have already been proposed in the literature to achieve this. \cite{mizrak2009detecting} can detect packet dropping or gray hole attacks in networks by exploiting the
correlation between packet delays and packet losses due to congestion. Their proposed methodology is based on passive observations of the one-way network delay experienced. For the scope of this work, the main advantage of this solution compared with the better-known proposals such as\cite{mizrak2009detecting} is that we could implement it without any additional overhead or support from the network -- \cite{mizrak2009detecting} assumes the routers in the network cooperate and provide real-time data related to the
queue lengths at their interfaces.

%%%%%%%%%%%%%%%%%%%%%%%%%%%%%%%%%%%%%%%%%%%%%%%%%%%%%%%%%%%%%%%%%%%%%%%%%%%
%START OF SECTION: Prevention Engine
%%%%%%%%%%%%%%%%%%%%%%%%%%%%%%%%%%%%%%%%%%%%%%%%%%%%%%%%%%%%%%%%%%%%%%%%%%%

\section{The Response Engine} \label{sec:response}
WedgeTail can be programmed to automatically reply to identified threats using its response engine. The response engine takes as input a set of XML-formatted policies and translates them into actions for the controller. Developing a fully fledged policy engine and ensuring the logical correctness of them is out of scope for this work. We developed a simplified policy engine for our initial evaluation of WedgeTail.  \par
\textbf{Policy Syntax:} Each policy requires six main features and attributes describing them. The features include Subject, Object, Actions, Condition, Exception and Validity time. Table 1 lists the attributes currently supported for these features. The naming used for attributes are assumed to be self-descriptive. Note that the values in parenthesis are expected to be provided as input for each of these attributes. While each policy may have only one subject, the other features may have more than just one associated attribute. The Exception attribute is mainly used to build hierarchy for the policies and validity is used to specify timing. \par
 We now look at two examples.  Let us revisit Figure \ref{Figure1:keyidea} and assume only FD(f) is detected as malicious. An administrator-defined policy may specify two different policies matching this forwarding device (i.e. one using the Forwarding Device attribute and another using Controller attribute). First, it may specify FD(g) as subject and instruct it to use an alternative route to forward traffic. Second, it may specify for the Controller to block all incoming OpenFlow messages. Now, consider the same scenario as before but this time with only FD(b) identified as malicious. In this case, there may be an Exception feature stating if a policy for FD(f) is still active then no action is executed from this policy.

\begin{table}
\centering
\begin{tabu} {|r|>{\raggedright}p{5.5cm}|p{2cm}|}
 \hline
 Feature & Attributes \\[2pt]
  \hline
Subject & Forwarding Device(id) | Controller  \\[2pt]
  \hline
Object & Packet(id) | Flow(id) | Switch(id)    \\[2pt]
  \hline
Action & Isolate(FD(id)) | Update\_forwarding\_table(FD(id)) | Alarm | Block\_Messages(FD(id)) | Test\_Again(FD(id)) \\
  \hline
Exception & Policy $P_i$ \\[2pt]
  \hline
Validity &  t (millisecond) \\
  \hline
\end{tabu}
\captionsetup{font=scriptsize}
\caption{Overview of the response engine policy syntax}
\end{table}
%We call the former, inter-plane trajectories and the latter, intra-plane trajectories.
%The maliciousness of a forwarding device can be identified by comparing the actual and expected trajectories . 

%\subsection{Compound Attacks}
%A compromised forwarding device may perform two or more of the aforementioned attacks for the same packet  in respect to one snapshots taken fro3m the network. For example, a $packet_i$ may be replayed and also 

%%%%%%%%%%%%%%%%%%%%%%%%%%%%%%%%%%%%%%%%%%%%%%%%%%%%%%%%%%%%%%%%%%%%%%%%%%%
%START OF SECTION: Prototype Implementation
%%%%%%%%%%%%%%%%%%%%%%%%%%%%%%%%%%%%%%%%%%%%%%%%%%%%%%%%%%%%%%%%%%%%%%%%%%%

\section{Implementation} \label{sec:implementation}
We envision WedgeTail to be integrated as an application for SDN controllers for both detection and response. However, at this stage, to demonstrate WedgeTail's compatibility with different platforms, evaluate it over different controllers and to ease the implementation we implemented the detection engine as a proxy sitting in between the control and data plane -- a similar architecture is also used in \cite{dhawan2015sphinx}. In fact, the detection engine requires advanced functions that, currently, is not consistently available across controllers.  Currently, the response engine is programmed as an application for Floodlight controller. WedgeTail's current architecture is shown in Figure \ref{Figure4}. \par
We implemented our system, mainly, in Java using approximately 1500 lines of code. WedgeTail work starts by creating scanning regions. To do this, it creates a unique hash from a large number of packets. The packets are then continuously tracked as they traverse the network by intercepting the PACKET\_IN messages sent from the data plane to control plane. This information is then used to create database records that list all the forwarding devices visited by packets along with some packet information and a timestamp attached to it. WedgeTail composes these records together and generates the actual packet trajectories using its Actual\_Trajectory\_Extractor module. \par
Once the scanning zones are generated, and the target forwarding devices identified, WedgeTailed requires having the expected trajectories of packets to initiate its inspection. Hence, WedgeTail queries the controller for current topology and launches a Mininet matching the same setup. It then intercepts all the OpenFlow messages exchanged between the control plane and data plane including FLOW\_MOD and PACKET\_IN messages. The OpenFlow messages sent from the controller to forwarding devices (e.g. FLOW\_MOD) is first translated into a database INSERT command. This command stores the rule, forwarding device receiving the rule along with a time value in a MySQL database. Thereafter, using the DB\_to\_Mininet\_Translator component, these are retrieved from the database and translated into appropriate Mininet commands. The result is a virtual network replica, which is continuously updated. The virtual network replica is used by the Hassel\_Expected\_Trajectory module to compute the expected trajectories of packets\footnote{We acknowledge that the authors of\cite{pelekis2010unsupervised} provided the source code of their program, which we used in our Target Identification module.}.

%%%%%%%%%%%%%%%%%%%%%%%%%%%%%%%%%%%%%%%%%%%%%%%%%%%%%%%%%%%%%%%%%%%%%%%%%%%
%START OF SECTION: Evaluation
%%%%%%%%%%%%%%%%%%%%%%%%%%%%%%%%%%%%%%%%%%%%%%%%%%%%%%%%%%%%%%%%%%%%%%%%%%%

\section{Evaluation} \label{sec:evaluation}
\begin{table}
\centering
\begin{tabu} to 0.44\textwidth { | X[c] | X[c] | X[c] | X[c] | }
 \hline
 Number of & AARNet Setup & Zib54 Setup & Sprint Setup \\
  \hline
FD & 12   &  54 &  316 \\
  \hline
Subnet & 40   &  800 & 48,966  \\
  \hline
Rules & 391 & 21,387  & 15,649,486  \\
  \hline
Trajectory & 403 &  38,654 &  638,271 \\
  \hline
\end{tabu}
\captionsetup{font=scriptsize}
\caption{Overview of simulated networks}
\end{table}
We evaluated WedgeTail over simulated networks, which were different in terms of the number of forwarding devices, forwarding rules, network subnets, and trajectories -- with our latest simulation closely resembling real-world network conditions. We replicated a number of attacks against SDN networks that were previously reported in the literature and analyzed the accuracy of WedgeTail in detecting these attacks. In order to further evaluate WedgeTail's detection engine, we then wrote scripts that synthetically implanted a total 500 attacks covering all of the malicious actions specified in \S\ref{sec:threatmodel} in our simulated networks. \par
Here, we report on WedgeTail's accuracy and performance including its detection and prevention success, average detection time, user perceived latencies, overheads related to policy verification, etc. To further challenge WedgeTail's detection engine, we also introduced congestion to the simulated networks causing packet losses and measured the resulting false alarms. Moreover, we also compared our proposal with related works and argued how WedgeTail, in most cases, outperforms them both in detection and response. 
Finally, given that target identification and virtual network replica reconstruction are new features introduced as part of WedgeTail and may be of use in other domains, we report on their performance separately.

\subsection{Experimental Setup} \label{sub:networksetup}

We simulated three different networks namely \textit{AARNet Setup}, \textit{Zib54 Setup} and \textit{Sprint Setup}. AARNet Setup was used in our initial feasibility study and resembled a minimalistic backbone ISP network topology with only 12 forwarding devices. The forwarding rules in this network reached 390, and we generated benign traffic such that about 400 trajectories existed in the snapshots taken for inspection by WedgeTail. In Zib54 Setup, we extended our simulated network to more than 50 forwarding devices. Snapshots are taken for analysis contained up to 21000 forwarding rules and 38000 trajectories over 800 subnets.
A large network is presumed to have more forwarding devices as well as many more trajectories at any given instance. These motivated us to evaluate WedgeTail under Sprint Setup. The Sprint Setup contained 316 forwarding devices with more than 600000 trajectories over a network with more than 15 million rules and 48000 subnets.

\textbf{Network Topologies:} The network topologies for AARNet Setup, Zib54 and Sprint were extracted from The Internet Topology Zoo \cite{knight2011internet}, SNDlib\cite{orlowski2010sndlib} and Rocketfuel \cite{spring2004measuring}, respectively. Figure 11.a and 11.b (in Appendix) represent AARNet Setup and Zib54. In these setups, each node is assumed to contain only one forwarding device, and there is only one link in-between these devices as also depicted in the figures. Figure 11.c (in Appendix) depicts the interconnection of different domains at Sprint backbone network, which we used as the network topology for Sprint Setup. Note that in Figure 11.c, for clarity, the forwarding devices at each node are not depicted, and only one link connects the nodes to each other.

\textbf{Flow Entries:} We were unaware of any publicly available flow entry data set for our simulated networks. Hence, to add flow-entries, we created an interface for a \textit{subset of} prefix found in a full BGP table from Route Views \cite{routeview} and spread them randomly and uniformly to each router as `local prefixes'. We then computed forwarding tables using shortest path routing. The resulting forwarding rules and subnets for each setup are shown in Table I. We report that a similar methodology is also adopted by relevant work such as \cite{chao2016securing}, \cite{zeng2014libra}.

\textbf{Traffic Generation:}
We used Mausezahn \cite{Mausezahn} and a custom script to add benign traffic to the networks. Similar to \cite{dhawan2015sphinx}, our custom written script imported three real-world network traces from \cite{3, 4, 7} to drive traffic into Mininet.

We hosted the simulated networks on a machine equipped with Intel Core i5, 2.66 GHz quad-core CPU and 16 GB of RAM. The SDN controller equipped with WedgeTail was hosted on a machine with Intel Core i7, 2.66 GHz quad-core CPU and 8 GB of RAM. 
\subsection{Attack Scenarios} \label{sub:attackscenarios}
We replicated all the attacks presented by the authors of \cite{dhawan2015sphinx} against networks using ODL, Floodlight, POX and Maestro as controllers. We also implemented further attacks including Network-to-Host DoS attack. All of these attacks use one or more of the capabilities defined in \S\ref{sec:threatmodel} and were successfully detected by WedgeTail.
As examples, we briefly revise the main characteristics of six different attacks and discuss how WedgeTail successfully detects all of them. We also compare the advantages of WedgeTail with SPHINX when detecting these threats -- note that the authors did not provide the source of their solution and therefore, we cannot provide a numerical performance comparison at this stage. We report on the performance metrics separately in \S9.4 and \S9.5. \par
\textbf{I. Network DoS:} In this case, compromised forwarding devices direct traffic into a loop and magnify a flow until it completely fills out the available link bandwidth. We report that all four controllers were vulnerable to this attack and this completed in sub-second time intervals. \par
\textbf{DETECTION:} This attack involves a compromised forwarding device that either generates, misroutes or replays packet(s). These anomalies can be easily detected using the trajectory-based attack detection algorithms presented in \S\ref{sec:detection}. Compared to SPHINX, WedgeTail does not rely on any administrator defined policies for detection of a Network DoS attack. \par
\textbf{II. Network-to-Host DoS:} Here, one or more forwarding devices send a large amount of traffic to the host network causing a DoS. This may bring down a host machine in extreme cases, and when dealing with mission critical systems, the impact would be catastrophic. Existing controllers do not have any detection mechanism against this attack. \par
\textbf{DETECTION:} Malicious forwarding device(s) may generate, replay or misroute packets towards a network host to cause a DoS attack. The result of the aforementioned actions is having unexpected trajectories in the network, which are automatically detected by WedgeTail. However, unless there are administrator-defined policies for each host, SPHINX is unable to detect Network-to-Host DoS. Furthermore, the number of policies to be processed in real-time will be a factor of the total number of hosts and forwarding devices. The performance of SPHINX when processing such large number of policies is unknown. Moreover, even with such policies in place, the attack may go undetected as the downlink to host may not reach any suspicious threshold (note that in most cases this attack adds a negligible processing overhead to the compromised forwarding device(s) and may also have a negligible impact on the bandwidth). 

\textbf{III. TCAM Exhaustion:} TCAM is a fast associative memory used to store flow rules. Malicious hosts may send arbitrary traffic and force the controller into installing a large number of flow rules, thereby exhausting the switch's TCAM. As also discussed in \ref{sec:threatmodel}, this may result in significant latency or packet drops. None of the controllers tested can detect nor prevent attacks such as TCAM Exhaustion. \par

\textbf{DETECTION:} Attacks similar to III results in packet delay or drop, which will result in anomalies between expected and actual trajectories and are detected by WedgeTail. SPHINX has a totally different approach for detecting such attacks. The latter checks for $FLOW\_MOD$ messages sent by the controller and detects a threat if the rate continues to be high over time. While both approaches will lead to detecting the threat, with SPHINX, the controller messages may not violate the administrator defined policies and still cause the switch to fail (e.g. the switch may be already experiencing a load higher than usual that is not covered in the policy description). In such cases, the attack will not be detected by SPHINX.

\textbf{IV. Forwarding device Blackhole:} In this case, flow path ends abruptly, and the traffic cannot be routed to the destination. A forwarding device either drops or delays packet forwarding to launch this attack. We installed malicious rules on switches in networks, and none of the controllers had any mechanism to prevent nor detect them. \par

\textbf{V. ARP Poisoning:} Malicious network hosts can spoof physical hosts by forging ARP requests and fool the controller to install malicious flow rules to divert traffic. This may be used for eavesdropping or in other cases to mount IP slicing attacks and creating network loops. We replicated the attack with the exact similar setup used in \cite{dhawan2015sphinx} and we also report that all of the tested controllers are vulnerable to it. Note that ARP poisoning corrupts the physical topological state. We discuss how WedgeTail detects attacks targeting the logical topological state in \S\ref{sec:discussion}. \par

\textbf{DETECTION:} There are no network policies that a forged ARP request violates in a network. However, the actual path that a packet traveling from hosts to the controller takes is visible to WedgeTail. Hence, ARP requests with an anomalous trajectory (i.e. originated from hosts rather than forwarding devices) can be monitored and blocked before poisoning the network. SPHINX is also capable of detecting this attack either using its flow graph feature (which binds MAC-IP) or using administrator defined detection policies.

\textbf{VI. Controller DoS:} With OpenFlow, a packet that does not match any of the currently installed flow rules of a forwarding device is buffered, and an associated OFPT PACKET\_IN message containing the data packet's header fields is forwarded to the controller. When a controller receives a large number of new packet flows within a short period, its buffer is filled up and has to forward complete packets to the controller. This causes heavy computational load on the controller, and it may bring it down altogether. We used Cbench \cite{cbench} and flooded the controller with a high throughput of $PACKET\_IN$ messages to analyze the controllers' performance. Similar to \cite{dhawan2015sphinx}, we report that all except Floodlight exhibited this attack. However, while Floodlight throttles the incoming OpenFlow messages from switches as a prevention mechanism, the connection of the switches with the controller is broken when a large number of switches attempt to connect with it. \par
\textbf{DETECTION:} A compromised forwarding devices may execute this attack by either replaying packets or generating packets destined to the controller. If there are an abnormal number of trajectories between a forwarding device and a controller in a snapshot taken from the network, then WedgeTail will detect a threat and can react as per the policies defined by its administrator -- note that WedgeTail may compute the threshold number of trajectories over time period $\Delta\tau$ by itself or, the administrator could custom define this. SPHINX detects a controller DoS by observing the flow-level metadata and computing the rate of PACKET\_IN messages, which is compared with the administrator-defined policies. Compared to SPHINX, WedgeTail also has the advantage of computing the aggregated flow heading to the controller rather than each individual link.
\subsection{Attack Implantations} \label{sub:attackimplant}

As mentioned, WedgeTail successfully detected all of the attacks implemented in \S\ref{sub:attackscenarios}. However, to cover all of the malicious actions specified in \S\ref{sec:threatmodel} and perform extended performance analysis, we wrote scripts to implant 500 synthetic malicious threats in our simulated networks. The resulting malicious forwarding devices maliciously processed:
\textbf{1.} All packets on all ports in approx. 30\% of all attacks, \textbf{2.} A subset of packets on a specific port in approx. 19\% of all attacks, \textbf{3.} A subset of packets on a specific port in approx. 19\% of all attacks, \textbf{4.} Packets pertaining to a specific port in approx. 25\% of all attacks, \textbf{5.} A subset of packets pertaining to a specific port in approx. 15\% of all attacks, \textbf{6.} Packets destined to the control plane in approx. 11\% of all attacks. \par

\textbf{Malicious Actions:} We used custom scripts to randomly introduce synthetic malicious forwarding devices in our networks. The resulting forwarding devices maliciously replayed packets (40\% of all attacks), dropped packets (30\%), misrouted packets (5\%), generated packets (10\%), and delayed packets (15\%). A packet replay may be used in a range of threats (e.g. surveillance, DDoS, etc.) and is less likely to be detected compared to packet drops -- i.e. traffic not reaching the destination is presumably much more noticeable. Hence, this distribution of attacks is deemed to be reasonable. \par

\textbf{Compound Attacks:} We define compound attacks as those involving more than one malicious forwarding device. For example, a surveillance attack may involve more than one malicious forwarding device (see Figure 1). Compound attacks are challenging for solutions such as SPHINX as compromised forwarding devices may intelligently install custom rules and avoid reporting to the controller thus aiming to conceal their maliciousness. This is less of an issue for WedgeTail's detection engine as any custom rule not matching those set by the control plane will eventually result in deviation of actual trajectories from expected ones, and this will trigger an alarm. We report that in our simulations a total of 108 attacks involved more than one malicious forwarding devices. Specifically, 35\% of these involved four malicious forwarding devices, 25\% six forwarding device, 40\% nine forwarding devices. In real-world scenarios, an attacker who has taken over nine forwarding devices of a network is a strong adversary. Specifically, in AARNet Setup this means that the 75\% of forwarding devices are compromised (this is a condition not supported by \cite{dhawan2015sphinx} requiring the majority of forwarding devices to be non-malicious). 

\subsection{Accuracy \& Detection Time}

We measured WedgeTail's detection accuracy in respect to A) Successful detection rate against attacks implanted in our simulated networks, B) Successful detection rate under network congestion leading to packet loss C) Successful application of pre-defined policies against malicious forwarding devices. \par 

For A, we implanted attacks as specified in \S\ref{sub:attackimplant} over our simulated networks. We then used WedgeTail to measure the absolute time taken to detect the faults. The detection time is defined as the time taken to raise an alarm from the instant a malicious packet is routed over the network by a forwarding device. We report that all of the 500 attacks implanted in the networks were successfully detected by WedgeTail. The distribution of attacks over the networks was as following: 50 were over AARNet Setup, 250 over Zib54 Setup and 200 over Sprint Setup. Essentially, AARNet Setup was part of our feasibility study stage and Zib54, and Sprint Setups serve to prove the practicality of WedgeTail in real-world. We illustrate the detection time of 50 attacks separately over network AARNet Setup, B and C in Figure \ref{fig:detectionsetupa}, \ref{fig:detectionsetupb} and \ref{fig:detectionsetupc}, respectively. The average detection time over AARNet Setup is about 54 second with a standard deviation of 12 seconds. For Zib54 Setup, the average detection time is about 705 second with a standard deviation of 80 seconds. For Sprint Setup, the average detection time is 5600 second with a standard 730 second. Moreover, the average detection times were not affected in the presence of Compound Attacks (see \S9.3). The latter is, in fact, expected given that the detection algorithm entails analyzing each and every forwarding device and the response engine is not triggered until the end of a full scan. \par
The aforementioned performance metrics show that WedgeTail's detection time scales well as the network size increases. The detection time of attacks over network snapshots is also acceptable. In other words, for a network administrator to be able to detect and locate malicious forwarding device after about 90 minutes without defining any policies or manual investigation is quite satisfactory. \par

For B, we added random congestion to the simulated networks, which resulted in packet drops at various points in the simulated networks. The dropped rate varied as 0, 0.005, 0.0075, 0.01, 0.015 and 0.02 of the 1K TCP flows sent over the simulated networks. Table 2 shows the overall detection results after detection delays of 3, 5 and 10 minutes -- WedgeTail attack detection is started after the detection delay time. Note that we added multiple bottlenecks throughout the networks. The results prove that packet loss due to congestion is not a prohibitive factor for our system. WedgeTail is now only able to distinguish between packet drops due to congestion and maliciousness. We acknowledge that we have not measured the impact of congestion on successful malicious forwarding device localization and we leave further investigation for our future work.
%Therefore, in our tests we did not assume congestion would add delay to packets traversing the network. As mentioned in \S\ref{sub:practicalconsiderations}, this could be addressed by using more advanced estimation techniques, and we leave this for future work. \par

Regarding C, we report that WedgeTail has matched policies with the threats and applied the actions specified in the policies for all attacks.

\begin{table}
\centering
\begin{tabu} to 0.44\textwidth { | X[c] | X[c] | X[c] | X[c] | }
 \hline
 Detection Delay & Accuracy & False Positive & False Negative \\
  \hline
3 minutes & 98.83  & 3 &  0.76 \\
  \hline
5 minutes & 99.17   & 3 & 0.69 \\
  \hline
10 minutes & 99.38 & 8  & 0.48  \\
  \hline
\end{tabu}
\captionsetup{font=scriptsize}
\caption{Overall detection results of attacks in the presence of packet drops due to congestion.}
\end{table}

\subsection{Performance Analysis}
In this section, we report on some of the main performance metrics of WedgeTail. Thereafter, we compare WedgeTail's performance with related work. \par
\begin{figure*}[!htb]
\minipage{0.32\textwidth}
\includegraphics[width=\linewidth]{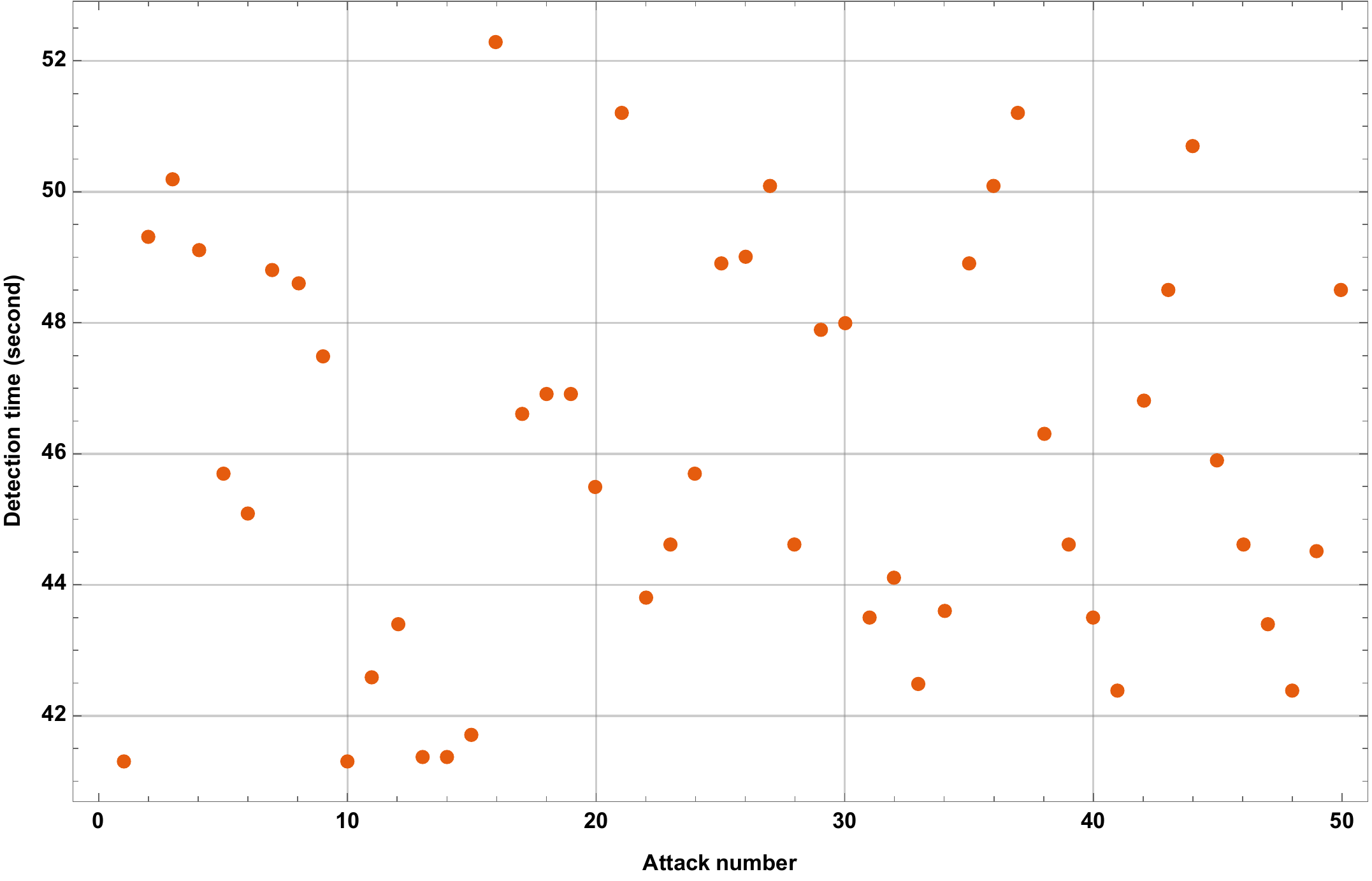}
  \caption{Attack detection in AARNet Setup (50 attacks).}\label{fig:detectionsetupa}
\endminipage\hfill
\minipage{0.32\textwidth}
\includegraphics[width=\linewidth]{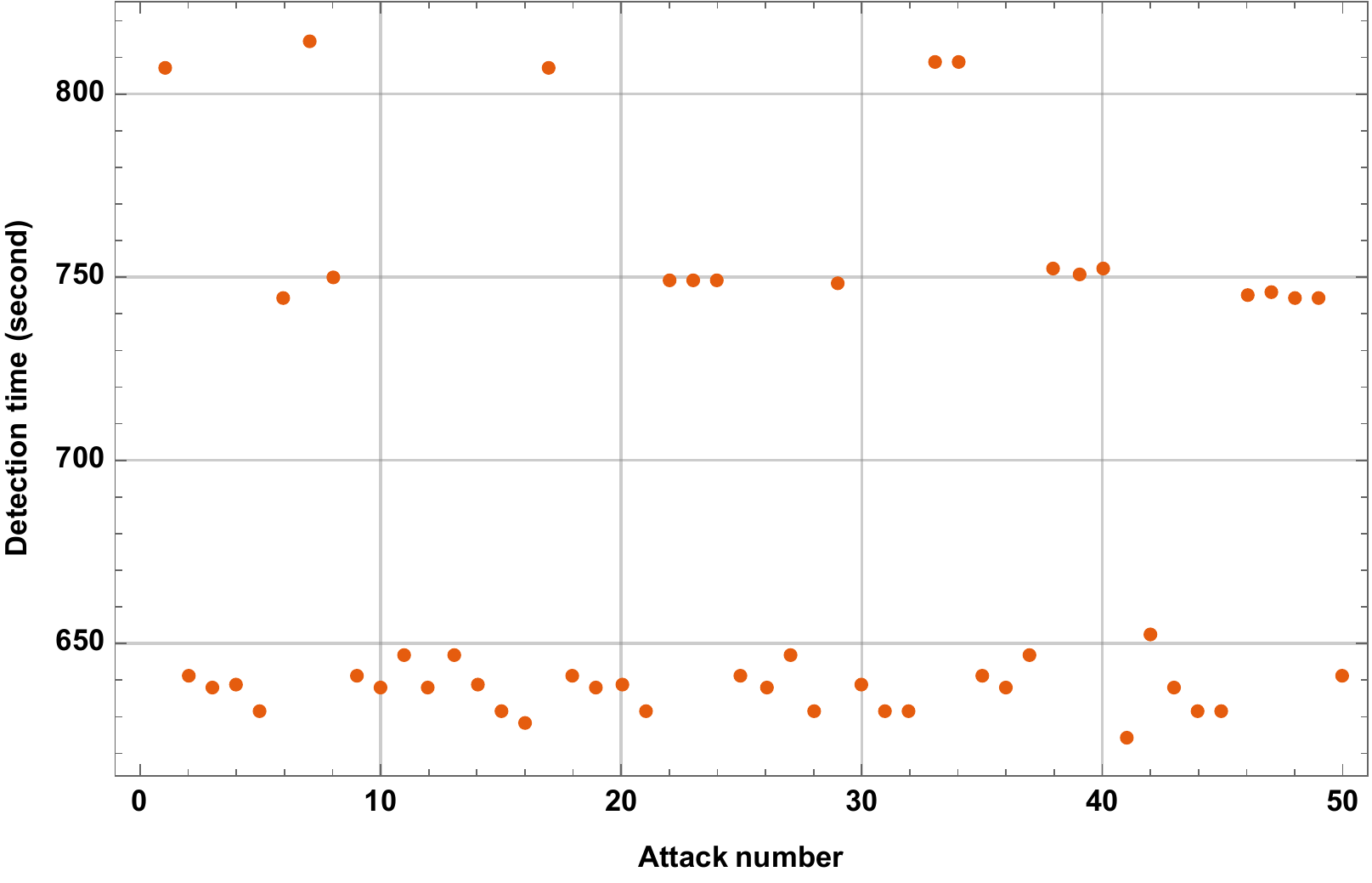}
  \caption{Attack detection in Zib54 Setup (50 attacks).}
\label{fig:detectionsetupc}
\endminipage\hfill
\minipage{0.32\textwidth}%
\includegraphics[width=\linewidth]{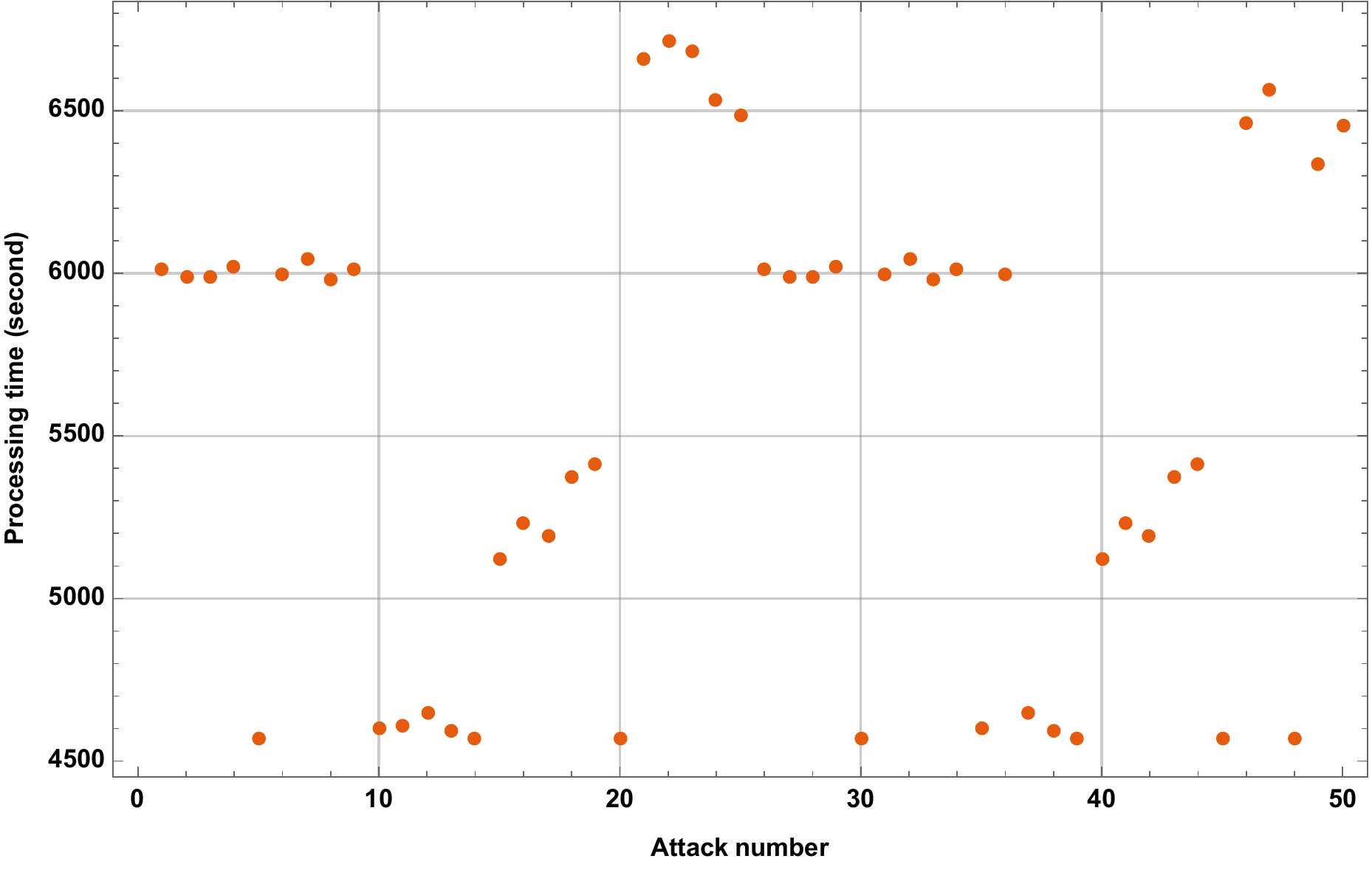}
  \caption{Attack detection in Sprint Setup (50 attacks).}
\label{fig:detectionsetupb}
\endminipage
\newline
\minipage{0.32\textwidth}
\includegraphics[width=\linewidth]{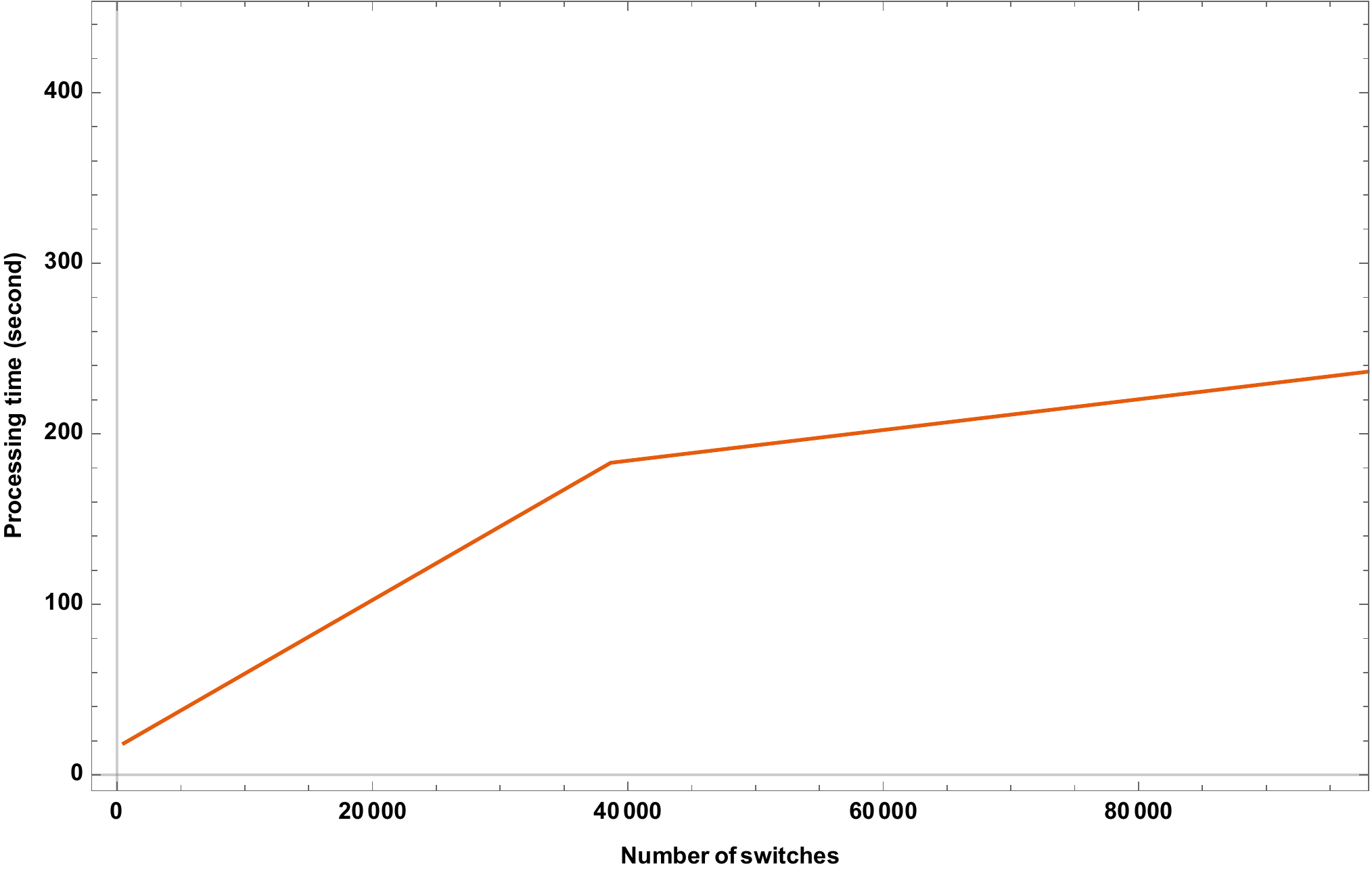}
  \caption{Target identification with respect to the number of forwarding devices.}\label{fig:targetidentificationfd}
\endminipage\hfill
\minipage{0.32\textwidth}
\includegraphics[width=\linewidth]{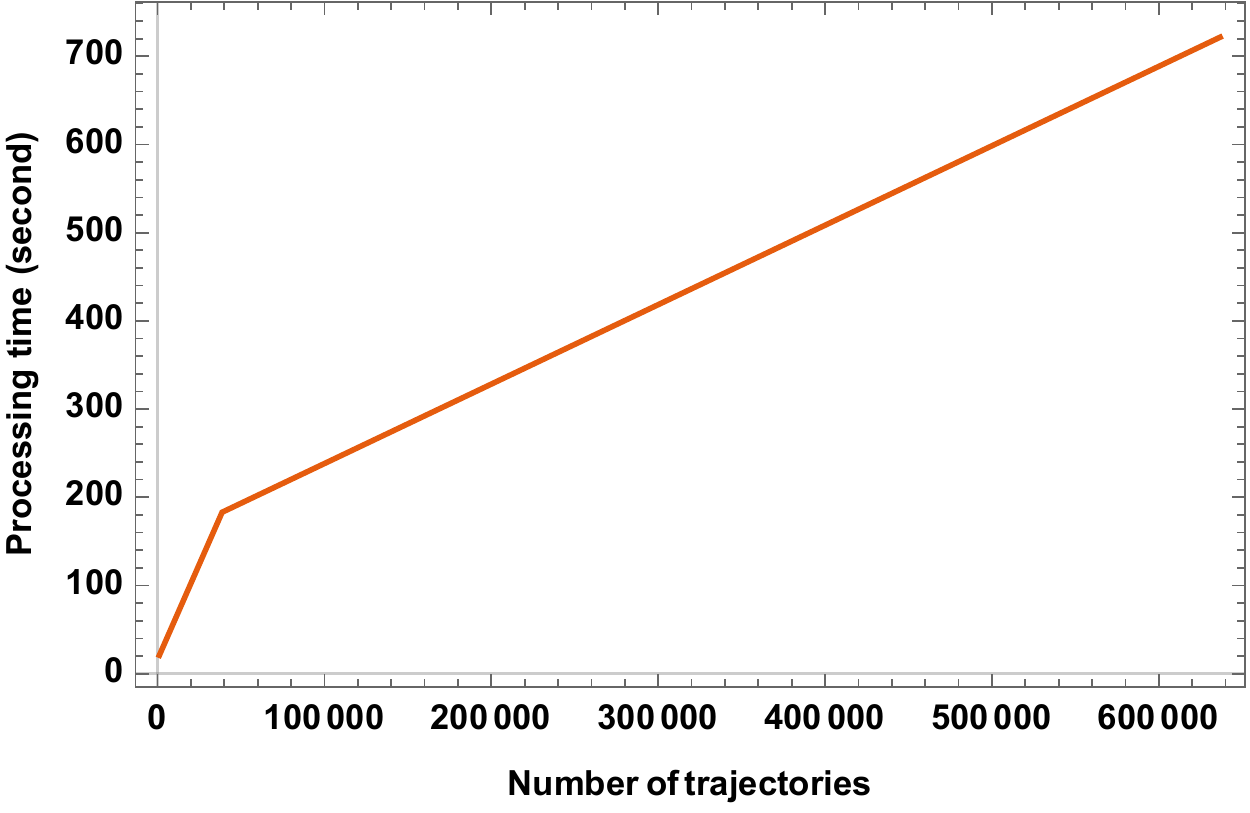}
  \caption{Target identification with respect to the number of trajectories.}\label{fig:targetidentificationtrajectories}
\endminipage\hfill
\minipage{0.32\textwidth}
\includegraphics[width=\linewidth]{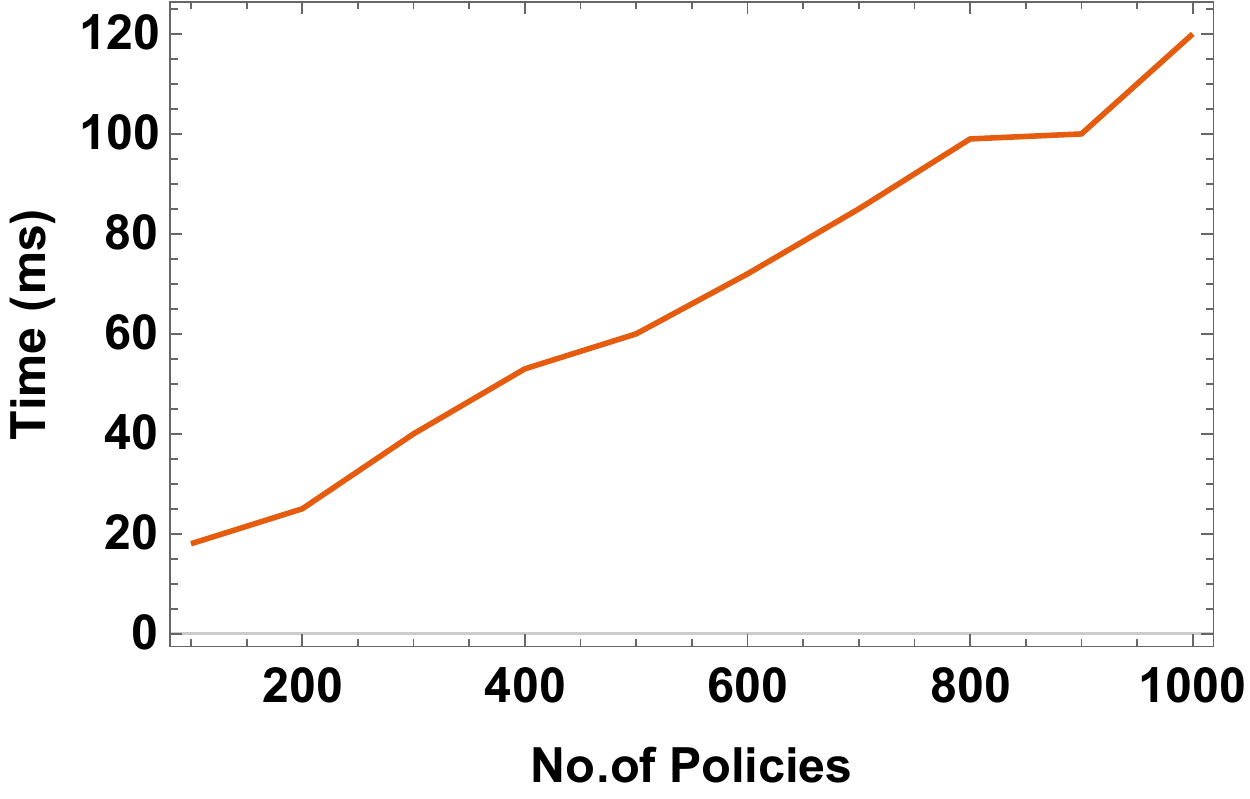}
  \caption{Average policy matching times with increasing policies.}\label{fig:policyverification}
\endminipage\hfill
\end{figure*}

\textbf{1. Target Identification:}
Figure \ref{fig:targetidentificationfd} illustrated the target identification time with respect to the number of forwarding devices that exist in networks. The algorithm takes approx. 18 seconds to identify the target forwarding devices in AARNet Setup with 400 trajectories. This number increases to up to 12 minutes for Sprint Setup, where there are approx. 640000 trajectories.

\textbf{2. Network Replica:} We calculated the replication delays after 500 instances of updates in the original network, and we observed an upper bound of approx. 15 seconds. To the best of our knowledge, this is the first system to maintain a virtual network replica both of the control plane and data plane in SDNs.  \par

\textbf{3. Response Policy Matching:} As shown in Figure \ref{fig:policyverification}, we observe that the average policy matching time as we increase WedgeTail's administrator defined policies from 0 to 1000 is approximately 120 milliseconds. Note that unlike SPHINX, WedgeTail's policies are used by its response engine only. 

\textbf{4. Resource Utilization:} We observe WedgeTail reaches a maximum CPU usage of approx. 15\% and memory usage of approx. 18\%. The CPU usage is mainly associated with target identification and packet tracking components of WedgeTail. 

\textbf{5. User Perceived Latencies:} WedgeTail is not a real-time system, and it has no implication for the network users when detecting threats. Comparatively, however, SPHINX adds overhead to the network and causes delays. Given the various advantages of WedgeTail compared to SPHINX in detection and prevention, we consider this a bonus feature for our system. \par

\textbf{Comparison with Related Work:} We discuss the reasons as to why WedgeTail is non-comparable to network troubleshooting solutions in \S\ref{sec:discussion}. However, to put WedgeTail's performance into perspective we report on the performance metrics of Anteater \cite{mai2011debugging}, which takes a snapshot of forwarding tables and analyze them for errors, and NetPlumber \cite{kazemian2013real} that extends HSA into a real-time verification solution. Anteater has been tested on a 178 router topology and takes more than 5 minutes to just check for loops. NetPlumber may take up to 10 minutes to verify network correctness after a given rule change. Comparatively, WedgeTail investigates for every instance of malicious action and does more than just evaluating rule sets (e.g. identifying scanning targets, tracking packets as they traverse the network, maintaining a network replica to remove trust from forwarding devices, etc.)  with a reasonably added overhead.

%\subsection{Performance Analysis} \label{sub:performance}
%As mentioned, the initial port and the set of target ports are defined by WedgeTail's administrator. We carried out an analysis of network traffic at one of the AARNET's node, the University of New South Wales (UNSW) for 20 days. We collected incoming and outgoing traffic three times a day resulting in about 10 GB traffic each day. The traces reveal that 9 different ports cover about 63\% of the traffic with most of the traffic pertaining to port 80 and 443\footnote{We are unable to list all of these ports as per the ethical agreement required for data collection.}. Hence, the initial port was set to HTTP throughout our tests. \par

%\begin{figure}
%  \centering
%	\includegraphics[height=4in, width=2.3in]{Figures/Distribution3}
%  \caption{a) the distribution of the attack scenarios, b) the distribution of the random synthetic malicious cases in the simulated networks.}
%  \label{Figure:distribution}
%\end{figure}

\section{The Good, the Bad and the Ugly} \label{sec:discussion}

\subsection{Why WedgeTail and SPHINX?} At this point, we would like to draw a clear line between network troubleshooting solutions and our work. Solutions such as \cite{khurshid2013veriflow, kazemian2013real, orlowski2010sndlib}, mainly, focus on the elimination of configuration conflicts, the avoidance of routing loops and black holes, the detection of policy inconsistency, and etc. However, even with a correct configuration, the forwarding devices may fail in execution due to bugs in switch software, conflicts, limited memory space. A simple failure to execution is itself worrisome but a malicious forwarding device is a serious threat to the network operators and associated hosts. Recently, threat from state actors and insiders are on the rise. With such strong adversaries, it is feasible to expect the attackers to exploit forwarding device vulnerabilities in the core of the networks to achieve their goals (e.g. surveillance, etc.). There are even simpler, and potentially more dangerous scenarios, when compromised forwarding devices are purposefully placed by insiders in such networks. Hence, we strongly believe networks require to have solutions built against such adversaries. One should note that these solutions should not be, mainly, measured regarding the detection time rather successful detection of all threats within a reasonable time. In summary, we regard WedgeTail as complementary to solutions provided for network troubleshooting and for this reason WedgeTail is built on top of the most robust proposals in the network troubleshooting domain including HSA\cite{kazemian2012header}.

\subsection{Why WedgeTail?}
Related works including \cite{dhawan2015sphinx}, mainly, rely on administrator-defined policies for attack detection, are built against weaker adversarial settings and fail to detect certain types of attacks (see \S\ref{sub:attackscenarios}). Moreover, they do not discuss localization of malicious forwarding devices, imposes some overhead to network performance, cannot distinguish between malicious actions such as packet drop or delay and do not prioritize the inspection of forwarding devices. 

\subsection{Limitations and What's Next?}

We evaluated WedgeTail over various network setups, configurations, and sizes equipped with different SDN controllers to prove its practicality under simulated environments closely matching real-world networks.  Specifically, WedgeTail's high accuracy and performance over Sprint Setup with a large number of forwarding devices, rules, and trajectories forms a solid ground motivating further development and evaluation of our proposed solution. Furthermore, we remind that WedgeTail's core detection and response techniques such as trajectory-creation, scanning methodology and inspection algorithms are platform independent and network dynamics do not alter these. Therefore, our next step is to deploy our solution over a real-world network setup focusing on scalability. \par
We also admit that we would need exploring WedgeTail's accuracy under more attack scenarios and use-cases (e.g. virtualization, VM migrations, and etc.). However, given our current evaluations results we do not expect any major hindrance for our steps forward. \par 
Another issue to point out is that our system analyzes snapshots and the stability of snapshots may be challenging \cite{zeng2014libra} -- as with all other similar offline systems proposed. Finally, WedgeTail's compatibility with distributed SDN controllers such as ONOS requires further investigation -- although we regard such platforms to be an enabler rather than a barrier. We aim to address these limitations in the near future.

\section{Conclusion} \label{sec:conclusion}
In the era of cyber-war, cyber-terrorism and with insider threats reportedly on the rise, it is to expect for attackers to exploit the vulnerabilities of the network core infrastructure to launch attacks against networks. Currently, Software Defined Networks (SDN) is regarded as the networks of the future. The SDN control plane security has been an ongoing topic of research. However, malicious forwarding devices could potentially be a more worrying threat as these are the actual enforcement point of decisions made at the control plane. Accordingly, SPHINX \cite{dhawan2015sphinx} was the first attempt in the literature to detect a broad class of attacks in SDNs with a threat model not requiring trusted switches or hosts. With the same set of goals, we proposed an alternative solution, which we call WedgeTail. Our solution is designed against stronger adversarial settings and outperforms prior solutions in various aspects including accuracy, performance, and autonomy.

%All the advantage that SDN could bring will not encourage widespread adoption if the security aspect of SDN networks is not thoroughly. In this paper, ...
%\textbf{[.. look at the threat model, discuss how WedgeTail covers it .. Discuss the challenges such as port picking and argue how it will assist in preventing threats .. highlight that this is the first IPS for this matter .. ]}
%\textbf{[how to make it a real-time solution, improve the port picking with some ideas ?! .. ]}
%This is, in fact, case-specific and dependent on the priority given by the controller to WedgeTail's provided input. 
%}
%netpubler ==> realtime.

%}
%It is important to note that given the static nature of HSA, the trajectories have to be built until the forwarding state of the network changes compared to when the snapshot is taken. Also, we remind that HSA is run according to the replicated configurations stored by WedgeTail. As we will discuss in \S\ref{sec:conclusion}, it is also possible to have this as a real-time solution if NetPlumber, an extension of HSA is used.
%%%%%%%%%%%%%%%%%%%%%%%%%%%%%%%%%%%%%%%%%%%%%%%%%%%%%%%%%%%%%%%%%%%%%%%%%%%
%START OF SECTION: Bibliography
%%%%%%%%%%%%%%%%%%%%%%%%%%%%%%%%%%%%%%%%%%%%%%%%%%%%%%%%%%%%%%%%%%%%%%%%%%%
\section*{Acknowledgments}
The authors would like to express their gratitude and appreciation to all the anonymous reviewers for their comments on the paper. Specifically, we are grateful to our Shepherd Dr. Cong Wang for his valuable feedback and assistance in improving the quality of this work. The first author also acknowledges the technical suggestions and recommendations of his former colleagues at Information Security Research Group of University College London (UCL).

\bibliographystyle{abbrv}
\small{
\bibliography{sigproc}}

\clearpage
\appendix
\section{Network Topologies}
For the sake of completeness, we include a representation of network topologies used in our evaluations in Figure \ref{fig:topology}. Figure 14a shows the topology used in AARNet Setup. The Image and topology for this are extracted from the Internet Topology Zoo \cite{knight2011internet}. Figure 14b illustrates the topology used in Zib54 Setup. The image as well as topology are extracted from SNDlib\cite{orlowski2010sndlib}. 
Figure 14c shows the network topology used in Sprint Setup. In this setup, each node of the figure is constituted of multiple interconnected forwarding devices. Image and topology are extracted from \cite{spring2004measuring}. \newline \newline \par

\begin{minipage}{\textwidth}
\begin{minipage}{.5\textwidth}
  \centering
  \includegraphics[height=160pt, width=160pt]{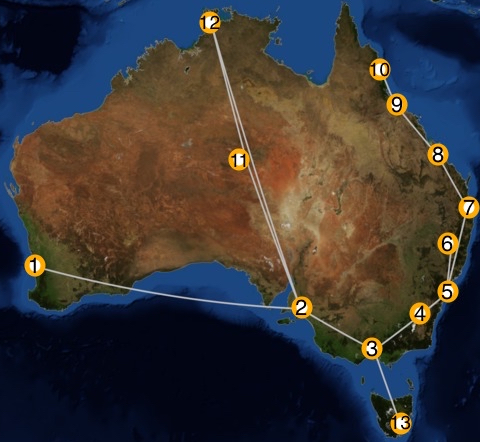}
  \captionof{subfigure}{AARNet network topology simulated in AARNet Setup.}
  \label{fig:aa}
\end{minipage}%
\begin{minipage}{.5\textwidth}
  \centering
  \includegraphics[height=160pt, width=180pt]{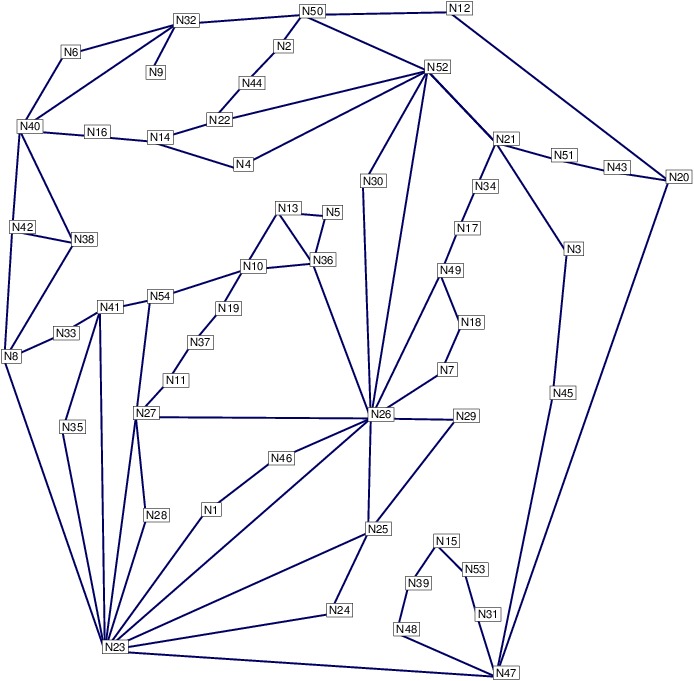}
  \captionof{subfigure}{Zib54 network topology simulated in Zib54 Setup.}
  \label{fig:b}
\end{minipage} \newline
\center
\begin{minipage}{.5\textwidth}
  \centering
  \includegraphics[height=130pt]{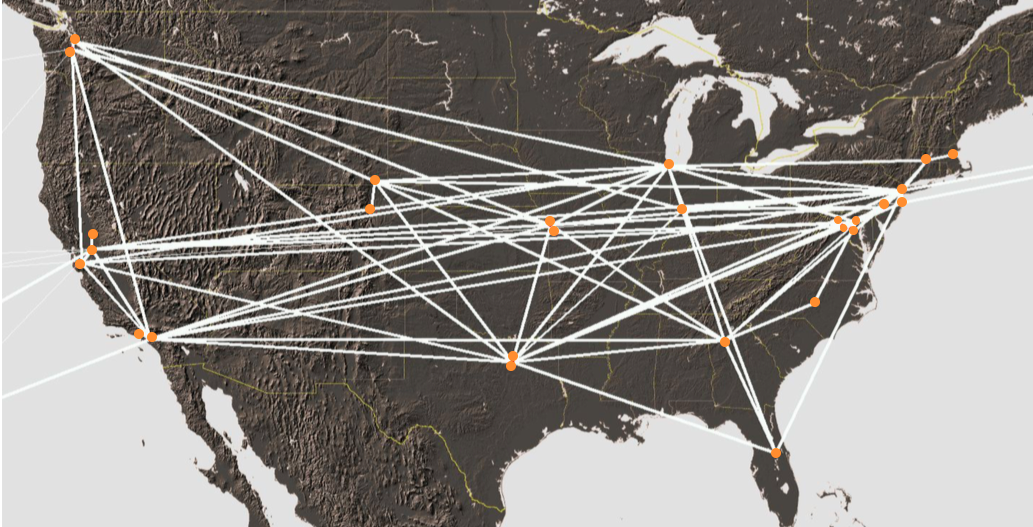}
  \captionof{subfigure}{Backbone topology of Sprint simulated in Sprint Setup.}
  \label{fig:c}
\end{minipage}%
\captionof{figure}{Network Topologies used in WedgeTail Evaluations}
\label{fig:topology}
\end{minipage}

\end{document}